\documentclass[aps,pr,twocolumn,showpacs,superscriptaddress,
groupedaddress,10pt]{revtex4}
\usepackage{amsmath}
\usepackage{amssymb}
\usepackage{graphicx}
\usepackage{epsfig}
\usepackage{dcolumn}
\usepackage{bm}
\usepackage{bm}
\usepackage{blindtext}
\usepackage{natbib}

\def\avg#1{\langle#1\rangle}
\def\Re{\rm{Re}}
\def\Im{\rm{Im}}
\def\be{\begin{equation}} \def\ee{\end{equation}}
\def\bea{\begin{eqnarray}} \def\eea{\end{eqnarray}}

\def\nn{\nonumber}
\def\pp{\parallel}

\begin{document}
\title{Topological septet pairing with spin-$\frac{3}{2}$
fermions -- high partial-wave channel counterpart of the $^3$He-B phase} 
\author{Wang Yang}
\affiliation{Department of Physics, University of California,
San Diego, California 92093, USA}
\author{Yi Li}
\affiliation{Princeton Center for Theoretical Science, Princeton 
University, Princeton, NJ 08544}
\author{Congjun Wu}
\affiliation{Department of Physics, University of California,
San Diego, California 92093, USA}

\begin{abstract}
We systematically generalize the exotic $^3$He-B phase, which not only 
exhibits unconventional symmetry but is also isotropic and 
topologically non-trivial, to arbitrary partial-wave channels 
with multi-component fermions.
The concrete example with four-component fermions is illustrated 
including the isotropic $f$, $p$ and $d$-wave pairings in the spin 
septet, triplet, and quintet channels, respectively.
The odd partial-wave channel pairings are topologically non-trivial,
while pairings in even partial-wave channels are topologically trivial.
The topological index reaches the largest value of $N^2$ in the $p$-wave
channel ($N$ is half of the fermion component number).
The surface spectra exhibit multiple linear  and even high order Dirac 
cones. 
Applications to multi-orbital condensed matter systems and
multi-component ultra-cold large spin fermion systems  are discussed.
\end{abstract}
\maketitle

Superconductivity and paired superfluidity of neutral fermions possessing
unconventional symmetries are among the central topics of condensed matter
physics.
If Cooper pairs formed by spin-$\frac{1}{2}$ fermions carry non-zero spin,
their orbital symmetries are in the odd partial-wave channels.
The $p$-wave paired superfluidity \cite{leggett1975,volovik2003} 
includes the $^3$He-A phase 
exhibiting point nodes \cite{anderson1961}, the fully gapped B phase
\cite{balian1963}, and the recently reported polar state with linear nodes 
\cite{aoyama2006,dmitriev2015}.
The $p$-wave superconductivity has also been
extensively investigated in SrRu$_2$O$_4$
\cite{mackenzie2003,nelson2004,kidwingira2006}, and 
heavy fermion systems including UGe$_2$, URhGe, UCoGe
\cite{aoki2012}.
The $p$-wave superfluid $^3$He and superconductors exhibit rich topological 
structures of vortices and spin textures under rotations or in 
external magnetic fields, respectively \cite{scharnberg1980,salomaa1987}.
In addition, experimental signatures of the possible nodal $f$-wave superconductivity have 
also been reported in UPt$_3$ \cite{machida2012,schemm2014}.

Among these unconventional pairing phases, the $^3$He-$B$ phase is 
distinct: in spite of its non-$s$-wave pairing symmetry and 
spin structure, the overall pairing structure remains isotropic 
and fully gapped.
Its pairing exhibits the relative spin-orbit symmetry breaking 
from $SO_L(3) \otimes 
SO_S(3)$ to $SO_J(3)$ \cite{leggett1975}
where $L$, $S$, and $J$ represent the orbital, spin, and
total angular momentum, respectively.
The relative spin-orbit symmetry-breaking has also been studied 
in the context of Pomeranchuk instability
termed as unconventional magnetism leading to dynamic generation
of spin-orbit coupling \cite{wu2004a,wu2007}.

Furthermore, the $^3$He-$B$ phase possesses non-trivial topological 
properties \cite{kitaev2001,schnyder2008,ryu2010}.
Topological states of matter have become a major research focus
since the discovery of the integer quantum Hall effect 
\cite{klitzing1980,kohmoto1985,thouless1982}.
Recently, the study of topological band structures has extended
from time-reversal (TR) breaking systems to TR invariant
systems \cite{kane2005,hasan2010,qi2011}, from two to three
dimensions \cite{schnyder2008,moore2007,roy2010}, 
and from insulators to superconductors \cite{kitaev2001,schnyder2008,ryu2010,
alicea2011,sau2010,lutchyn2010,qi2010}.
The $^3$He-B phase is a 3D TR invariant topological Cooper pairing state.
Its bulk Bogoliubov spectra are analogous to the 3D gapped Dirac fermions
belonging to the DIII class characterized by an integer-valued index 
\cite{schnyder2008}. 
The non-trivial bulk topology gives rise to the gapless surface
Dirac spectra of the mid-gap Andreev-Majorana modes \cite{chung2009}.
Evidence of these low energy states has been reported in 
recent experiments \cite{bunkov2015}.

Because the electron Cooper pair can only be either spin singlet or
triplet, the $p$-wave $^3$He-B phase looks the only choice of the 
unconventional 3D isotropic pairing state.
In this article, we will show that actually there are much richer
possibilities of this exotic class of pairing in all the partial-wave
channels of $L\ge 1$.
We consider multi-component fermions in both orbital-active solid 
state systems and ultra-cold atomic systems with large spin
alkali and alkaline-earth fermions,
both of which have recently attracted a great deal 
of attention  \cite{wu2012,ho1999,wu2003,wu2006,deSalvo2010,gorshkov2010,
taie2010,fang2015,fu2010}.
For simplicity, below we introduce an effective spin $s$
to describe the multi-component fermion systems with the component
number expressed as $2N=2s+1\ge 4$.
Compared with the 2-component case, their Cooper pair spin
structures are greatly enriched \cite{wu2010a,ho1999}.
For example, the 4-component spin-$\frac{3}{2}$ systems can support 
the $f$-wave septet, $p$-wave triplet, and $d$-wave quintet pairings, 
all of which are fully gapped and rotationally invariant.
Nevertheless, only the odd partial-wave channel ones, i.e., the
$p$ and $f$-wave pairings are topologically non-trivial.
Their topological properties are analyzed both from calculating
the bulk indices and surface Dirac cones of the Andreev-Majorana modes.
For the $p$-wave case, the topological indices from all the
helicity channels add up leading to a large value of $N^2$.
Correspondingly the surface spectra exhibit 
the coexistence of 2D Dirac cones of all the orders
from 1 to $2N-1$.

We begin with an $f$-wave spin septet Cooper pairing 
Hamiltonian in a 3D isotropic system of spin-$\frac{3}{2}$ 
fermions
\bea
H=\sum_{\vec k} \epsilon_{\vec k} c^\dagger_\alpha(\vec k) c_\alpha(\vec k)
-\frac{g}{V_0}\sum_{\vec k, \vec k^\prime,m,\nu} P^\dagger_{m,\nu}(\vec k)
 P_{m,\nu}(\vec k^\prime),
\label{eq:Ham}
\eea
in which $\epsilon_{\vec k}=\frac{\hbar^{2}k^2}{2m}-\mu$ and
$\mu$ is the chemical potential.
$\alpha =\pm\frac{3}{2},\pm\frac{1}{2}$ is the spin index,
$g$ is the pairing interaction strength, and $V_0$ is the system volume.
The pairing operator is defined as
$P^\dagger_{m,\nu}(\vec k)=c^\dagger_\alpha(\vec k) Y_{3m}(\hat k)
[S^{3\nu} R]_{\alpha\beta} c^\dagger_\beta(-\vec k)$ 
where $\hat k=\vec k/k$, $Y_{3m}(\hat k)$'s with $-3\le m\le 3$
are the 3rd order spherical harmonic functions, and
$S^{3\nu}$ with $-3\le \nu \le 3$ are the rank-3 spherical tensors 
based on the spin operator $\vec S$ in the spin $\frac{3}{2}$-representation,
where $\nu$ is the eigenvalue of $S_z$.
For later convenience,  $Y_{3m}(\hat k)$ are  normalized
according to $\sum_m |Y_{3m}(\hat k)|^2=1$.
$R$ is the charge conjugation matrix defined as
$R_{\alpha\beta}= (-)^{\alpha+\frac{1}{2}}\delta_{\alpha, -\beta}$
satisfying 
$R \vec S^T R^{-1}=-\vec S$ such that $R_{\alpha\beta}c^\dagger_\beta$
transforms in the same way under rotation as $c_\alpha$ does.
The expressions for spherical harmonic functions and spin tensors
are presented in Appendix \ref{sect:tensor}.

After the mean-field decomposition, Eq. \ref{eq:Ham} becomes
\bea
\frac{H_{MF}}{V}=\frac{1}{V}\sum^\prime_{\vec k}\Psi^\dagger(\vec k) 
H(\vec k) \Psi(\vec k) +g  \sum_{m,\nu}\Delta^*_{m,\nu} \Delta_{m,\nu},
\label{eq:mf}
\eea
in which $\vec k$ is summed over half of momentum space; 
$\Psi(\vec k)=(c_{\vec k,\alpha}, c^\dagger_{-\vec k,\alpha} )^{T}$
is the Nambu spinor; the order parameter $\Delta_{m,\nu}$ 
is defined through the self-consistent equation as
\bea
\Delta_{m,\nu}=\frac{g}{V}\sum_{\vec k}
\avg{G|c_\gamma(-\vec k) Y^*_{3m}(\vec k) R^\dagger S^{3\nu,\dagger} 
c_\delta(\vec k)|G}
\eea
with $\avg{G|...|G}$ meaning the ground state average.
The matrix kernel $H(\vec k)$ in Eq. \ref{eq:mf} is expressed as
\bea
H(\vec k)= \epsilon(\vec k)\tau_3 \otimes I_{4\times 4}
+\hat \Delta(\vec k)\tau_+ +\hat \Delta(-\vec k)\tau_-,
\label{eq:pairham}
\eea
where $\tau_3$ and $\tau_\pm= \frac{1}{2}( \tau_1\pm i \tau_2)$ are the
Pauli matrices acting in the Nambu space.
$\hat \Delta(\vec k)$ is defined in 
the matrix form in spin space as
\bea
\hat \Delta(\vec k)= \sum_\nu (S^{3\nu} R)  d^{*,\nu}(\vec k),
\label{eq:pair_matrix}
\eea
where $d^{*,\nu}(\vec k) = \Delta_{m,\nu} Y_{3m} (\hat k)$ and
is dubbed as the {\it $d$-tensor} in analogy to the $d$-vector in $^3$He.
The usual $d$-vector is represented in its three Cartesian
components, while here, the $d$-tensor is a rank-3 complex spherical tensor.

We consider the isotropic pairing with total angular 
momentum $J=0$, which is a generalization of the $p$-wave $^3$He-B phase.
Similarly, it is fully gapped, and thus conceivably  
energetically favorable within the mean-field theory.
Its $d^\nu(\vec k)$ can be parametrized as
$d^\nu(\vec k)=c_{f} \Delta_f (\frac{k}{k_f} )^3
Y_{3\nu}(\hat k)$, where $c_{f}$ is an overall normalization
factor given in Appendix \ref{sect:matrix}.
$\Delta_f$ is the complex gap magnitude, or, equivalently, 
\bea
\hat \Delta (\vec k)=\Delta_f (\frac{k}{k_f})^3 K_f(\hat k) R 
\label{eq:dvector}
\eea
in which $K_f = c_{f} U(\hat k) S^{30}U^\dagger(\hat k)$;  $U(\hat k)$
rotates the $z$-axis to $\hat k$ as
defined in the following gauge 
$U(\hat k)=e^{-i\phi_k  s_z} e^{-i\theta_k s_y}$ in which
$\theta_k$ and $\phi_k$ are polar and azimuthal angles of $\hat k$,
respectively. 
The explicit form of $\hat\Delta (\hat k)$ and the corresponding 
spontaneous symmetry breaking pattern are presented in
Appendices \ref{sect:matrix} and \ref{sect:symm},
respectively.

With the help of the helicity operator $h(\hat k)=\hat k \cdot \vec S$,
$K_f(\hat k)$ can be further expressed in an explicitly rotational 
invariant form as
\bea
K_f(\hat k)= -\frac{5}{2}  h^3(\hat k)
+\frac{41}{8} h(\hat k),
\label{eq:K}
\eea
which is diagonalized as $U^{\dagger}(\vec k) K_f(\hat k)U(\hat k)=
(-\frac{5}{2}S_{z}^{3}+\frac{41}{8}S_{z})$.
For a helicity eigenstate with the eigenvalue $\lambda$,
the corresponding eigenvalue $\xi_\lambda$ of $K_f(\hat k)$ 
reads $\xi_{\lambda}=-\frac{3}{4},\frac{9}{4},-\frac{9}{4}, \frac{3}{4}$
for $\lambda=\frac{3}{2},\frac{1}{2},-\frac{1}{2},-\frac{3}{2}$,
respectively. 
The Bogoliubov quasi-particle spectra are 
$E_{\lambda}(\vec k)=\sqrt{\epsilon^2(\vec k)+|\Delta_f|^2
(\frac{k}{k_f})^6\xi_{\lambda}^{2}}$ satisfying
$E_\lambda(\vec k)=E_{-\lambda}(\vec k)$ due to the parity 
symmetry.

\begin{figure}
\centering\epsfig{file=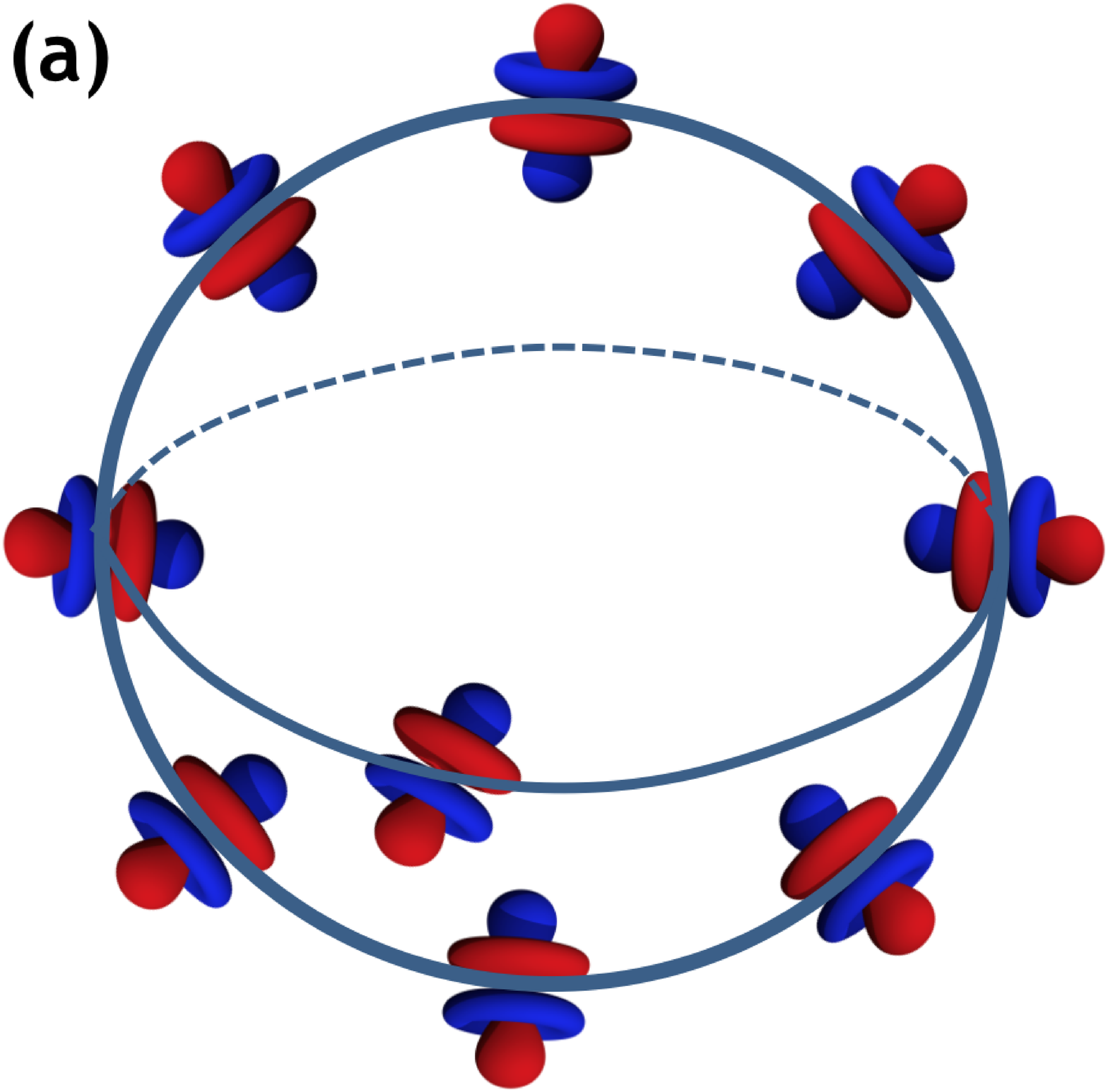,clip=1,width=0.49\linewidth,angle=0}
\centering\epsfig{file=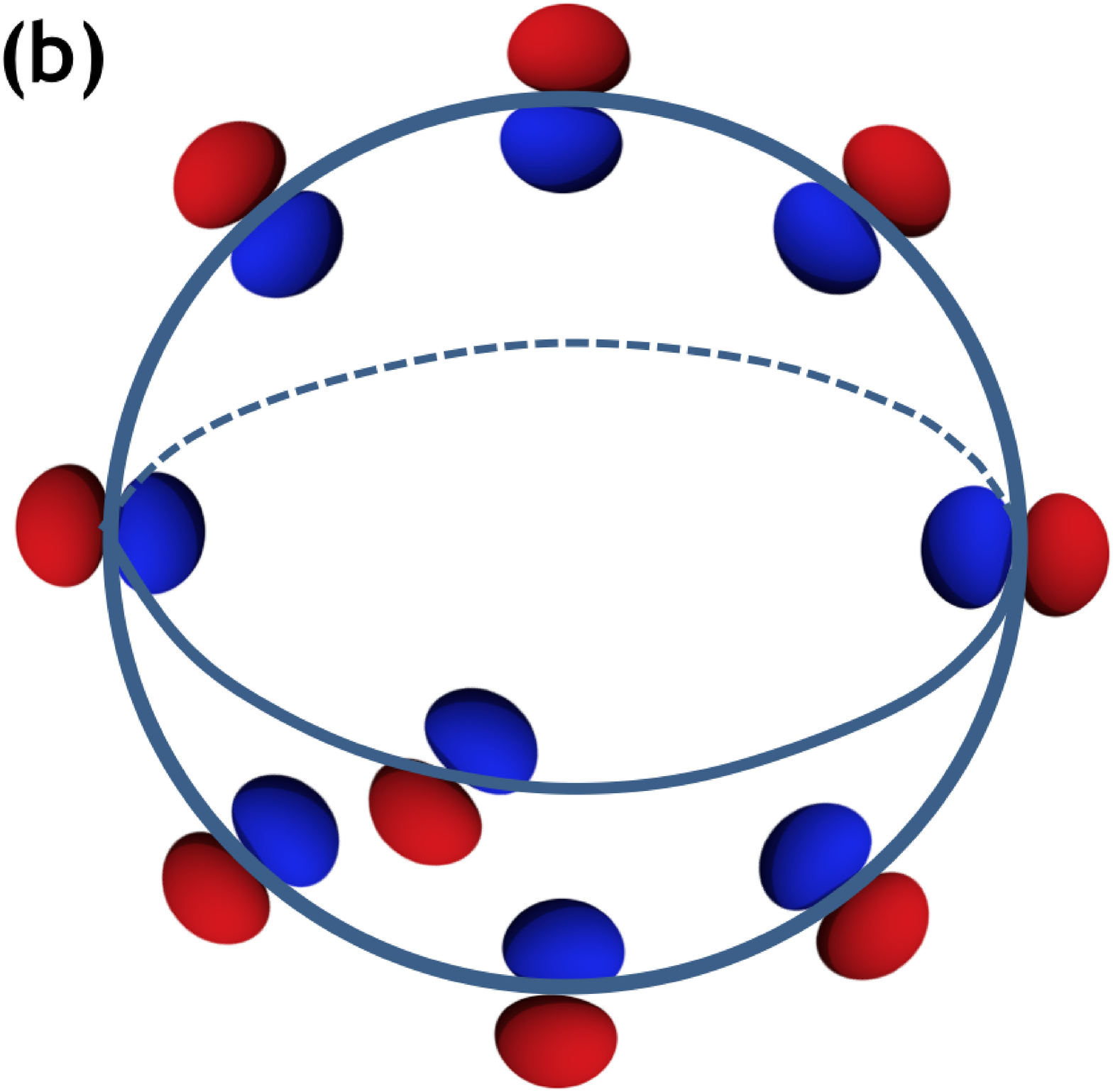,clip=1,width=0.49\linewidth,angle=0}
\caption{Pictorial representations of the pairing matrices over the Fermi 
surfaces of ($a$) the isotropic $f$-wave septet pairing  and ($b$) the 
isotropic $p$-wave triplet pairing with spin-$\frac{3}{2}$ fermions.
Intuitively, the $f$-wave matrix kernels $U(\hat k)S^{30} U^\dagger(\hat k)$ 
and the $p$-wave ones $U(\hat k)S^{10} U^\dagger(\hat k)$ for each 
wavevector $\vec k$ are depicted in their orbital counterpart harmonic 
functions in $(a)$ and $(b)$, respectively. 
}
\label{fig:order}
\end{figure}

Next we study the pairing topological structure. 
The pairing Hamiltonian Eq. \ref{eq:pairham} in the Bogoliubov-de Gennes
(B-deG) formalism possesses the particle-hole symmetry $C_p H(\vec k) C_p^{-1}=
-H^*(-\vec k)$ with $C_p=\tau_1\otimes I_4$.
Furthermore, the isotropic pairing state described by Eq. \ref{eq:dvector} 
is TR invariant satisfying $C_T H(\vec k) C_T^{-1}=H^*(-\vec k)$ 
with $C_T=I_2\otimes R$, and thus it belongs to the DIII class.
The associated topological index is integer-valued which will
be calculated following the method in Ref. [\onlinecite{qi2010}].
$H(\vec k)$ is transformed  with only two off-diagonal blocks 
as $\epsilon(\vec k) \tau_1 +\Delta_f (\frac{k}{k_f})^3 K(\hat k) \tau_2$.
The singular-value-decomposition to its up-right block yields 
$U(\hat k) L(k) \Lambda(k)U^\dagger(\hat k)$, in which
$L(k)$ and $\Lambda(k)$ are two diagonal matrices
only dependent on the magnitude of $k$ defined as 
$L_{\lambda\lambda}(k)=E_\lambda(k)$ and $\Lambda_{\lambda\lambda}(k)
=e^{i\theta_\lambda(k)}$, respectively.
The angles satisfy
$\tan \theta_\lambda(\vec k)= -\frac{\Delta_f \xi_\lambda} 
{\epsilon_{\vec k}} (\frac{k}{k_f})^3$ and for simplicity 
$\Delta_f$ is set as positive.
The $k^3$-dependence of the pairing amplitude is regularized:
Beyond a cutoff  $k_{c}$, $\Delta_f$ vanishes.

The topological index is calculated through
the SU(4) matrix $Q_{\vec k}=U(\hat k) \Lambda(k) U^\dagger(\hat k)$ as
\bea
N_{w}=\frac{1}{24\pi^2} \int d^{3}{\vec k}\epsilon^{ijl}
\mbox{Tr}[Q^{\dagger}_{\vec k}\partial_{i}Q_{\vec k}
Q^{\dagger}_{\vec k}\partial_{j}Q_{\vec k}
Q^{\dagger}_{\vec k} \partial_{l} Q_{\vec k}],
\label{eq:topoindex}
\eea
which is integer-valued characterizing the homotopic class of the 
mapping i.e., $\pi_{3}(SU(4))=\mathbb{Z}$.
Nevertheless, $N_w$ is only well-defined up to a sign:
After changing $\Delta_f\rightarrow -\Delta_f$, $N_w$ 
flips the sign.
As shown in Appendix \ref{sect:nw}, at $\mu>0$ $N_w$ is evaluated as
\bea
N_w=\sum_{\lambda=\pm\frac{3}{2},\pm\frac{1}{2}} \lambda ~\mbox{sgn}(\xi_\lambda).
\label{eq:nw}
\eea
Its dependence on $\mbox{sgn}(\xi_\lambda)$ is because
$\theta_\lambda(\vec k)$ varies from $0\rightarrow \frac{\pi}{2}
\rightarrow \pi$ at $\xi_\lambda>0$ but from
$\pi\rightarrow \frac{\pi}{2}\rightarrow 0$
as $k$ varies from $0$ to $k_f$ to $+\infty$.
A similar form of Eq. \ref{eq:nw} was obtained in Ref. 
[\onlinecite{qi2010}] in which the Fermi surface
Chern number plays the role of $\lambda$ in Eq. \ref{eq:nw}. 
For two helicity pairs of $\lambda=\pm\frac{3}{2}$ 
and $\lambda=\pm\frac{1}{2}$, their contributions are with
opposite signs, and thus $N_w=2$.

\begin{figure}
\includegraphics[height=0.49\columnwidth,width=0.48\linewidth]{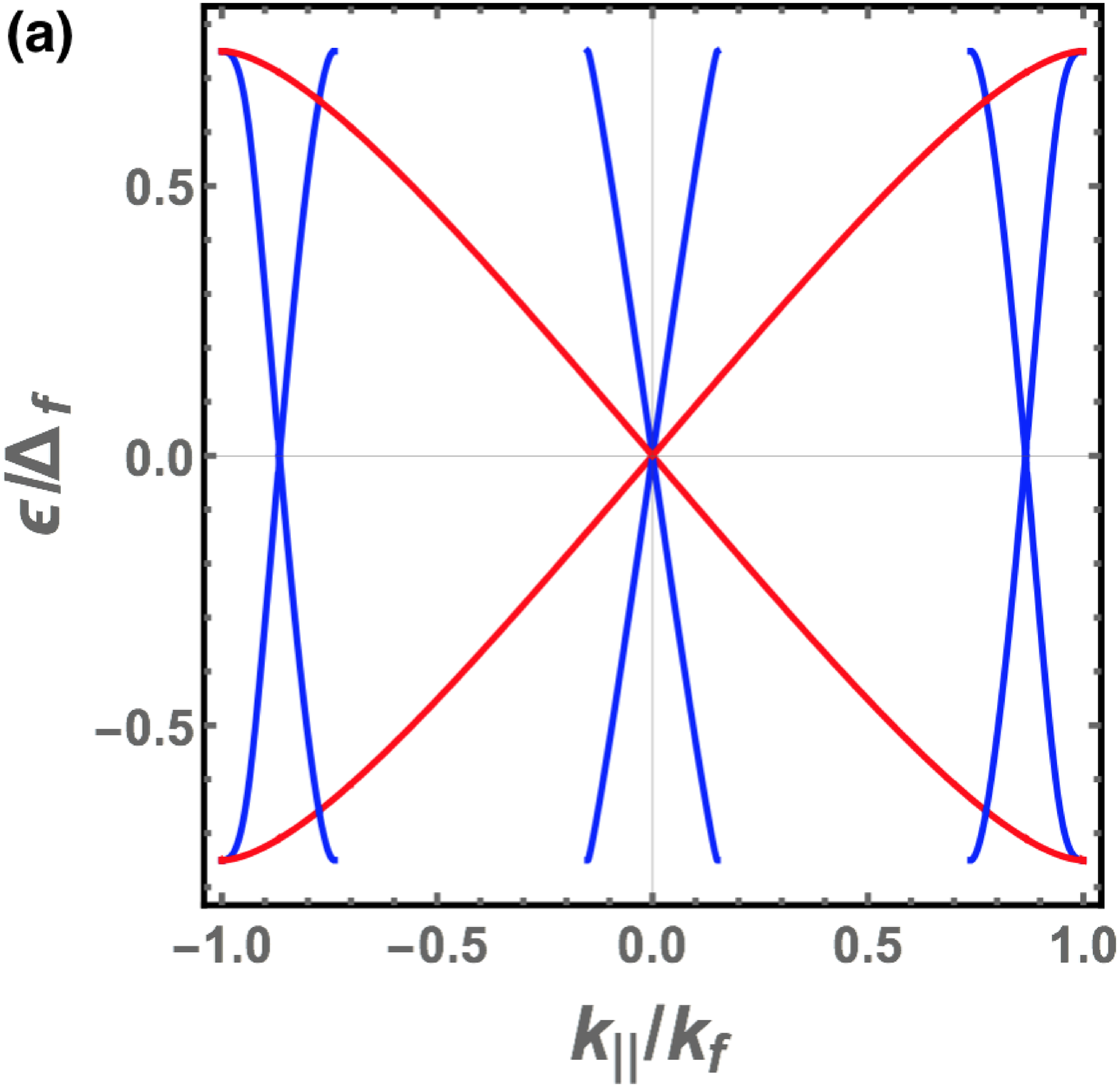}
\includegraphics[height=0.49\columnwidth,width=0.48\linewidth]{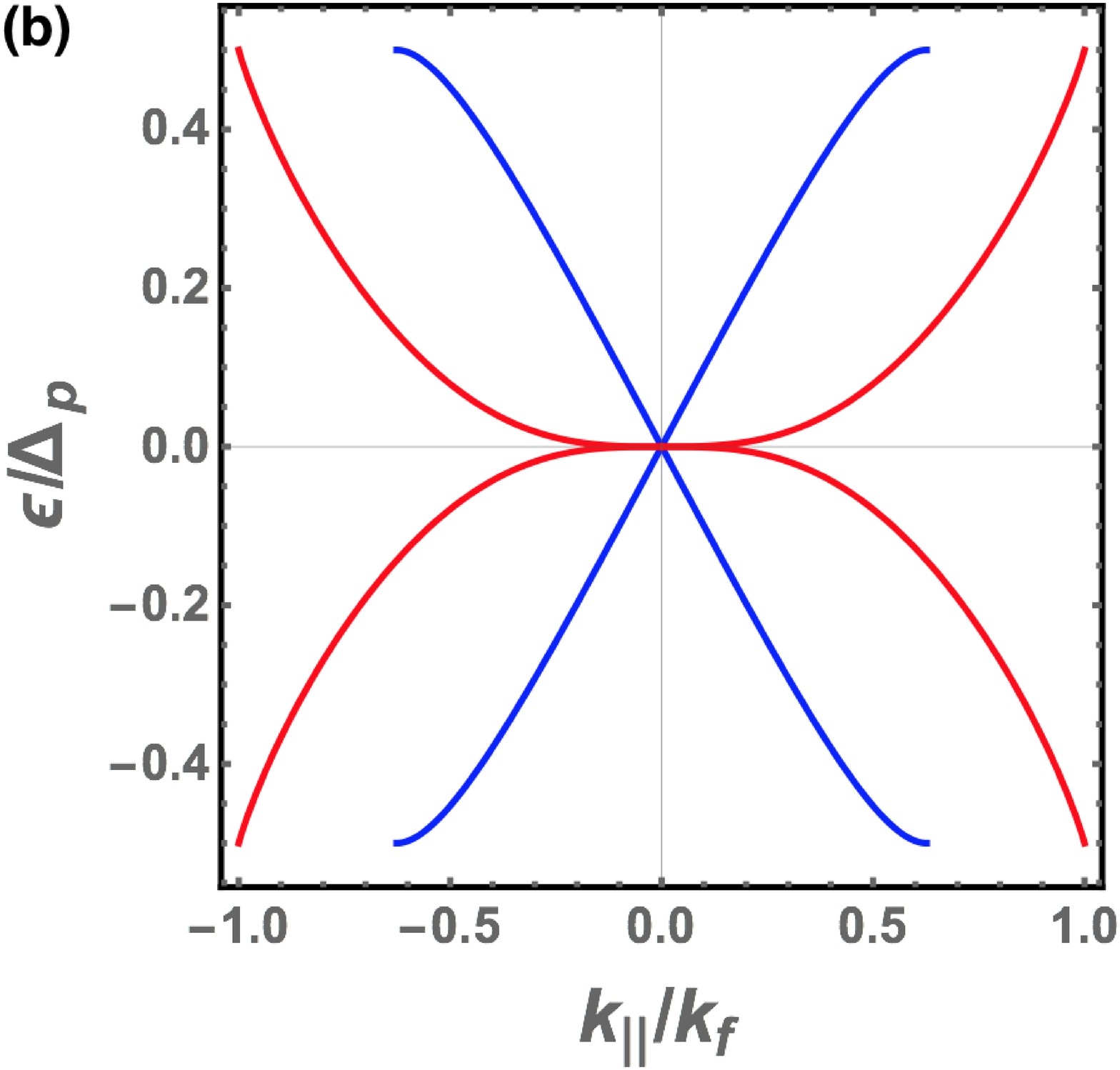}
\caption{The gapless surface spectra for the isotropic $f$-wave 
septet pairing in ($a$) and for the isotropic $p$-wave triplet
pairing in ($b$) with spin-$\frac{3}{2}$ fermions. 
}
\label{fig:surface}
\end{figure}

The non-trivial bulk topology gives rise to gapless surface Dirac cones.
Because of the pairing isotropy, without loss of 
generality, an open planar boundary is chosen at $z=0$
with $\mu(z)=\epsilon_f>0$ at $z<0$ and $-\infty$ at $z>0$.
The mean-field Hamiltonian becomes $H(\vec k_\parallel, z)$ in which 
$\vec k_\parallel=(k_x,k_y)$ remains conserved
while the translation symmetry along the $z$-axis is broken. 
The symmetry on the boundary is $C_{v\infty}$ including the uni-axial
rotation around the $z$-axis and the reflection with respect to any 
vertical plane.
$C_{v\infty}$ is also the little group symmetry at $\vec k_\pp=0$,
then the four zero Andreev-Majorana modes at $\vec k_\pp=0$ are 
$s_z$ eigenstates denoted as $|0_{\alpha,f}\rangle$.
The associated creation operators $\gamma^\dagger_{\alpha}$ are solved as
\bea
\gamma^\dagger_{\alpha}= \int^0_{-\infty} dz &[& e^{i(\frac{\varphi}{2}
+\frac{\pi}{4})}c^\dagger_{\alpha}(\vec k_\pp= 0,z) \nn \\
&+&e^{-i(\frac{\varphi}{2}+\frac{\pi}{4})}c_{-\alpha}(\vec k_\pp=0,z)]  u_{\alpha}(z),
\ \ \,
\label{eq:zero}
\eea
where $\varphi$ is the phase of $\Delta$; $u_{f,\alpha}(z)$ is the 
zero mode wavefunction exponentially decaying along the $z$-axis, 
and its expression is presented in Appendix \ref{sect:f_surface}.

The surface zero modes $|0_{\alpha,f}\rangle$ at $\vec k_\parallel=0$ possess 
an important property that the gapped bulk modes do not have:
They are chiral eigen-modes satisfying $C_{ch} |0_{\alpha,f}\rangle=
(-)^{\nu_\alpha} |0_{\alpha,f}\rangle$ with $\nu_\alpha=0$ for 
$\alpha=\frac{3}{2},-\frac{1}{2}$ and $\nu_\alpha=1$ for
$\alpha=\frac{1}{2},-\frac{3}{2}$,
respectively, in which
the chiral operator is defined as $C_{ch}= i C_p C_T=i\tau_1\otimes R$.
The mean-field Hamiltonian $H(\vec k_\pp, z)$ is in the DIII class satisfying
the particle-hole and TR symmetries, and it transforms as
$C_{ch} H(\vec k_\pp, z) C_{ch}^{-1}= -H(\vec k_\pp, z)$.
Thus $C_{ch}$ is a symmetry only for zero modes. 
For a nonzero mode $|\psi_n\rangle$ and its chiral partner 
$|\psi_{\bar n}\rangle=C_{ch}|\psi_n\rangle$, their energies are
opposite to each other, {\it i.e.}, $\epsilon_{\bar n}=-\epsilon_n$. 
If a perturbation $\delta H$ remains in the DIII class, then
$C_{ch} \delta H C_{ch}^{-1}= -\delta H$.
$\delta H$ can {\it only} mix two zero modes 
with opposite chiral indices because
$\avg{0_{\alpha,f}|\delta H|0_{\beta,f}}=(-)^{\nu_\alpha+\nu_\beta+1}
\avg{0_{\alpha,f}|\delta H|0_{\beta,f}}$, and it is nonzero only if
$\nu_\alpha\neq \nu_\beta$.



As moving away from $\vec k_\pp=0$, the zero modes evolve to the midgap 
states developing energy dispersions.
At $k_\pp \ll k_f$, these midgap states can be solved by using the 
$k\cdot p$ perturbation theory within the subspace spanned by the
zero modes $|0_{\alpha,f}\rangle$ at $\vec k_\pp=0$.
By setting $\delta H=H(\vec k_\pp,z)-H(0,z)$, the effective Hamiltonian to the
linear order of $k_\pp$ is
\bea
H_{mid}^f(\vec k_\parallel)=\frac{9 \Delta_f }{4 k_f} \left(
\begin{array}{cccc}
0&- ik_- & 0 & O (k_-^3 )\\
ik_+& 0 & -2i k_- &0 \\
0& 2 i k_+ &0 & -i k_- \\
O(k_+^3)&0&i k_+ &0
\end{array}
\right), \ \ \,
\label{eq:fmidgap}
\eea
where  $k_\pm=k_x\pm i k_y$.
The matrix elements in the same chiral sector are exactly zero,
and the elements at the order of $O(k^3_\pm)$ are neglected.
The solutions consist two sets of 2D surface Dirac cone spectra represented 
by $E^{a(b)}_{\pm} (\vec k_\parallel)=\pm v_{a(b)} k_\pp$.
The velocities are solved as $v_{a(b)}= \frac{9}{4} 
\frac{|\Delta_f|}{k_{f}} (\sqrt{2} \pm 1)$. 
We also develop a systematic method beyond the $k\cdot p$ theory 
to solve the midgap spectra for all the range of $k_\pp$ 
as presented in Appendix \ref{sect:general},
and the results are
plotted in Fig. \ref{fig:surface} ($a$).
In addition to the Dirac cones,
there also exists an additional zero energy ring not captured
by Eq. \ref{eq:fmidgap}, which is located at 
$k/k_f = \frac{\sqrt 3}{2}$ as analyzed
in Appendix \ref{sect:f_surface}.

Now we move to other unconventional isotropic pairings of 
spin-$\frac{3}{2}$ fermions in the $p$ and $d$-wave channels. 
The $p$-wave triplet one is topologically non-trivial, and 
the analysis can be performed in the same way as above.
The pairing matrix is
$
\hat\Delta_p(\vec k)=\sum_{\nu=0,\pm1} (S^{1\nu}R)d_p^{*,\nu}(\vec k)
=\Delta_p \frac{k}{k_f} K_p(\hat k) R,
$
where $S^{1\nu}$ is the rank-1 spin tensor, $d_p^{\nu}=\Delta_p
(\frac{k}{k_f})Y_{1\nu}(\hat k)$, and $K_p(\hat k)=\hat k \cdot \vec S$
is just the helicity operator.
The quasi-particle spectra are fully gapped as $E_\lambda(\vec k)
=\sqrt{\epsilon^2(\vec k)+|\Delta_p|^2(\frac{k}{k_f})^2 \lambda^2}$,
and the topological index of this pairing can be evaluated
based on Eq. \ref{eq:nw} by replacing the eigenvalues of
$K_f(\hat k)$ with those of $K_p(\hat k)$.
The contributions from two helicity pairs of 
$\lambda=\pm\frac{3}{2}$ and $\pm\frac{1}{2}$ add up leading to 
a high value $N_w=4$.
In comparison, the topological index of the $^3$He-B phase is only 1,
and thus their topological sectors are different
in spite of the same pairing symmetry.

The surface spectra of the isotropic $p$-wave pairing with 
spin-$\frac{3}{2}$ fermions are interesting: 
They exhibit a cubic Dirac cone in addition to a linear one.
Consider the same planar boundary configuration as before, 
similarly for each spin component $\alpha$ there exists one zero 
mode at $\vec k_\pp=0$ labeled by $|0_{\alpha,p}\rangle$.
Again we perform the  $k\cdot p$ analysis at $k_\pp\ll k_f$ in the
subspace spanned by $|0_{\alpha,p}\rangle$ with respect to 
$\delta H=H(\vec k_\pp,z)-H(0,z)$. 
The chiral eigenvalue of $|0_{\alpha,p}\rangle$  is 
$(-)^{\nu_\alpha}=\mbox{sgn}(\alpha)$, which leads to a different 
structure of effective Hamiltonian from that of the $f$-wave one.
Only $|0_{\pm\frac{1}{2},p}\rangle$ can be directly
coupled by $\delta H$, which leads
to a linear Dirac cone.
In contrast, the pair of states $|0_{\pm\frac{3}{2},p}\rangle$ are not
directly coupled, rather $|0_{\frac{3}{2},p}\rangle$ and 
$|0_{-\frac{1}{2},p}\rangle$ 
are coupled through the 2nd order perturbation theory, and 
so do $|0_{-\frac{3}{2},p}\rangle$ and $|0_{\frac{1}{2},p}\rangle$.
Consequently, $|0_{\pm\frac{3}{2},p}\rangle$ are coupled 
at the order of $(\delta H)^3$ developing a cubic Dirac cone
as shown in Appendix \ref{sect:p_surface}.
The above analysis is confirmed by the solution based on the 
non-perturbative method in Appendix \ref{sect:general}, 
as plotted in Fig. \ref{fig:surface} (b).

In contrast, the $d$-wave spin quintet isotropic pairing of
spin-$\frac{3}{2}$ fermions are topologically trivial. 
By imitating the analyses above, we replace $K_f(\hat k)$ with 
$K_d(\hat k)= 2 (\hat k\cdot \vec S)^2  - \frac{5}{2} I_{4} $.
Different from the kernels $K_p$ and $K_f$ in odd partial-wave channels,
$K_d$'s eigenvalues are even with respect to the helicity index, 
i.e., $\xi^d_\lambda=\xi^d_{-\lambda}$, such that $N_w$ vanishes.
This result agrees with that 3D TR invariant topological 
superconductors should be parity
odd as shown in Ref. [\onlinecite{zhangfan2013}].
The explicit calculation of the surface spectra in 
Appendix \ref{sect:d_surface} 
confirms this point showing the absence of zero modes.

The above analysis can be straightforwardly applied to multi-component 
fermion systems with a general spin value $s=N-\frac{1}{2}$.
The spin-tensors at the order of $l$ are denoted as $S^{lm}$ with 
$0 \le l\le 2S$ and $-l\le m \le l$. 
For each partial-wave channel $0\le l\le 2S $, there exists
an isotropic pairing  with the pairing
matrix $\hat \Delta(\hat k)=\Delta_l (\frac{k}{k_f})^l K_l(\hat k) R$ in 
which $K_l=U(\hat k ) S^{l0} U^\dagger(\hat k)$, whose
topological index $N_w(l)$ is determined by the sign pattern
of the elements of the diagonal matrix $S^{l0}$.
For even and odd values of $l$, $S^{l0}_{\alpha\alpha}=
\pm S^{l0}_{-\alpha-\alpha}$, respectively, and thus 
$N_w^l$ vanishes when $l$ is even, while
for odd values of $l$,
\bea
N_w(l)=\sum_{\lambda>0} 2\lambda ~\mbox{sgn}(S^{l0}_{\lambda\lambda}),
\eea
in which  $S^{l0}_{\alpha\alpha}=(-)^{\alpha+\frac{1}{2}} 
\avg{S\alpha, S-\alpha|SS;l0}$ up to an overall factor. 
The largest value of $N_w$ is reached for the $p$-wave case:
Since $S^{10}\propto S_z$, contributions from all the components 
add together leading to $N_w=N^2$.
The $^3$He-B phase of spin-$\frac{1}{2}$ fermions and the isotropic $p$-wave 
pairing with spin-$\frac{3}{2}$ fermions are two examples. 
As for the surface zero modes $|0_{\alpha,p}\rangle$ at $\vec k_\pp=0$,
their chiral indices equal $\mbox{sgn}(\alpha)$.
As a result, similar to the spin-$\frac{3}{2}$ case, when performing 
the $k\cdot p$ analysis for midgap states within the subspace
spanned by $|0_{\alpha,p}\rangle$, only $|0_{\pm\frac{1}{2},p}\rangle$ 
are directly coupled leading to a linear Dirac cone, and other 
pairs of $|0_{\pm\alpha,p}\rangle$ are indirectly coupled
at the order of $(\delta H)^{2\alpha}$ leading to 
high order Dirac cones.

Multi-component fermion systems are not rare in nature.
In solid state systems, many materials are orbital-active including
semiconductors, transition metal oxides, and heavy fermion systems.
Due to spin-orbit coupling, their band structures are denoted
by electron total angular momentum $j$ and in many situations
$j>\frac{1}{2}$.
For example, in the hole-doped semiconductors, the valence band carries 
$j=\frac{3}{2}$ as described by the Luttinger model \cite{winkler2003}.
Superconductivity has been discovered in these systems including 
hole-doped diamond and Germanium \cite{takano2004,ekimov2004,
herrmannsdorfer2009}.
Although in these materials, the Cooper pairings are mostly 
of the conventional $s$-wave symmetry arising from the electron-phonon
interaction, it is natural to further consider
unconventional pairing states in systems with similar band
structures but stronger correlation effects. 
The $p$-wave pairing based on the Luttinger model has 
been studied in Ref. [\onlinecite{fang2015}]. 
In ultra-cold atom systems,
many alkali and alkaline-earth fermions often carry large 
hyperfine spin values $F>\frac{1}{2}$, and thus their Cooper pair
spin structures are enriched taking values from $0$ to $2F$
not just singlet and triplet as in the spin-$\frac{1}{2}$ case
\cite{ho1999,wu2006,wu2010a}.

In multi-component solid state systems, there often exists
spin-orbit coupling.
For example, the Luttinger model describing
hole-doped semi-conductors \cite{winkler2003}, contains an isotropic 
spin-orbit coupling $H_{so}=\gamma_2 k^2 (\hat k \cdot \vec S)^2$.
Since $H_{so}$ is diagonalized in the helicity eigenbasis, we only need
to update the kinetic energy with $\epsilon_{k\lambda}=\epsilon_k +\gamma k^2
\lambda^2$ in the mean-field analysis, which satisfies
$\epsilon_{k\lambda}=\epsilon_{k,-\lambda}$, 
and the pairing structure described by 
Eq. \ref{eq:dvector} is not affected.
The topological properties are the same as analyzed before because 
the index formula Eq. \ref{eq:nw} remains valid and the surface 
mid-gap state calculation can be performed qualitatively similarly.
Nevertheless, the symmetry breaking pattern is changed.
The relative spin-orbit symmetry is already explicitly broken by the $H_{so}$.
The spin-orbit coupled Goldstone modes in $^3$He-B become gapped 
pseudo-Goldstone modes with the gap proportional to
the spin-orbit coupling strength $\gamma_2$.

In summary, we have found that multi-component fermion systems can support 
a class of exotic isotropic pairing states analogous to the $^3$He-B phase 
with unconventional pairing symmetries and non-trivial topological structures. 
High-rank spin tensors are entangled with orbital partial-waves at the 
same order to form isotropic gap functions.
For the spin-$\frac{3}{2}$ case, the odd partial-wave channel pairings 
carry topological indices 2 and 4 for the $f$ and $p$-wave pairings, 
respectively, while the $d$-wave channel pairing is topologically trivial.
The surface Dirac cones of mid-gap modes are solved analytically which 
exhibit two linear Dirac cones in the $f$-wave case, and the coexistence 
of linear and cubic Dirac cones in the $p$-wave case. 
Generalizations to systems with even more fermion components
can be performed straightforwardly.
This work provides an important guidance to search for 
novel non-trivial topological pairing states in both
condensed matter and ultra-cold atom systems.

{\it Acknowledgments}
W. Y. and C. W. are supported by the NSF DMR-1410375 and AFOSR FA9550-14-1-0168.
Y. L. is grateful for the support from the Princeton Center for
Theoretical Science.
C. W. acknowledges the supports from the National Natural Science Foundation 
of China (11328403), the CAS/SAFEA International Partnership Program 
for Creative Research Teams of China, and the President's Research 
Catalyst Awards CA-15-327861 from the University of California
Office of the President.

{\it Note added.}
After the submission of this manuscript, the evidence for the septet 
pairing with spin-$\frac{3}{2}$ fermions has been reported in the 
rare earth-based half-Heusler superconductors \cite{kim2016}.


\appendix
\section{Spherical harmonic functions and high-rank Spin tensor operators}
\label{sect:tensor}

In this section, we present spherical harmonic functions in momentum
space and high-rank spin tensors.

For convenience in the main text, we normalize the spherical harmonic functions 
$Y_{lm}(\hat k)$ defined on the Fermi surface satisfying 
\bea
\sum_{m=-l}^l |Y_{lm}(\hat k)|^2=1.
\eea
This normalization differs from the usual one of $\int d\Omega_k 
|Y_{lm}(\hat k)|^2=1$ only by an overall factor 
$\sqrt{\frac{4\pi}{2l+1}}$.
More explicitly, for the $p$-wave case, they are defined as
\bea
k Y_{1\pm1}(\hat k)&=&\mp\frac{1}{\sqrt 2}k_\pm, \ \ \, \ \ \,
k Y_{10}(\hat k)= k_z, 
\eea
For the $d$-wave case, they are defined as
\bea
k^2 Y_{2\pm2}(\hat k)&=&\sqrt{\frac{3}{8}} k_\pm^2, \ \ \, \ \ \,
k^2 Y_{2\pm1}(\hat k)=\mp\sqrt{\frac{3}{2}} k_\pm k_z, \nn \\
k^2 Y_{20}(\hat k)&=&\frac{1}{2} (3k_z^2-k^2).
\eea
For the $f$-wave case, they are defined as
\bea
k^3 Y_{3\pm3}(\hat k)&=& \mp\frac{\sqrt 5}{4} k_\pm^3, \ \ \, \ \ \,
k^3 Y_{3\pm2}(\hat k)= \frac{\sqrt {30}}{4}k_\pm^2 k_z, \nn \\
k^3 Y_{3\pm1}(\hat k)&=& \pm\frac{\sqrt 3}{4} k_\pm(k^2-5 k_z^2), \nn \\
k^3 Y_{30}(\hat k)&=& - \frac{1}{2} (3k^2-5k_z^2) k_z.
\eea
All of them are homogeneous polynomials of momentum components
$k_x$, $k_y$ and $k_z$.

The spin-$\frac{3}{2}$ matrices are defined in the standard way as
\bea
S_{+}&=&\left(\begin{array}{cccc}
0& \sqrt{3}&0 & 0\\
0 & 0 & 2 & 0\\
0& 0 & 0 & \sqrt{3}\\
0& 0 &0&0
\end{array}\right), \nn \\
S_-&=&S_+^\dagger,\nn \\
S_{z}&=&\left(\begin{array}{cccc}
\frac{3}{2}&0&0&0\\
0&\frac{1}{2}&0&0\\
0&0&-\frac{1}{2}&0\\
0&0&0&-\frac{3}{2}
\end{array}\right),
\eea
in which $S_\pm=S_x\pm i S_y$.
The general rank-$k$ spin tensors $S_{jm}$ satisfy
\bea
[S_-, S_{jm}]&=&\sqrt{(j+m)(j-m+1)} S_{jm-1}.
\eea
Based on these relations, we can build up spin tensors.
For example, the rank-1 tensors are defined as
\bea
S_{11}=-\frac{1}{\sqrt 2}S_+, \ \ \,
S_{10}=S_z, \ \ \, S_{1-1}=\frac{1}{\sqrt 2} S_-.
\eea

The rank-2 tensors are spin quadrupole operators defined as
\bea
S_{22}&=&\frac{1}{\sqrt 3}S_{11}^2=\left(\begin{array}{cccc}
0&0&1&0\\
0&0&0&1\\
0&0&0&0\\
0&0&0&0
\end{array}
\right)\nn \\
S_{21}&=&\frac{1}{2}[S_-, S_{22}]=\left(\begin{array}{cccc}
0&-1&0&0\\
0&0 &0&0\\
0&0 &0&1\\
0&0 &0&0
\end{array}
\right) \nn \\
S_{20}&=&\frac{1}{\sqrt 6}[S_-, S_{21}] =\frac{1}{\sqrt 2}
\left(\begin{array}{cccc}
1&0&0&0\\
0&-1&0&0\\
0&0&-1&0\\
0&0&0&1
\end{array}
\right)
\nn \\
S_{2-1}&=&\frac{1}{\sqrt 6}[S_-,S_{20} ]=-S_{21}^\dagger\nn \\
S_{2-2}&=&\frac{1}{2}[S_-,S_{2-1}]=S_{22}^\dagger.
\eea
This set of tensors can be organized into the Dirac $\Gamma$ 
matrices through the relations of
\bea
\Gamma_1&=&-i (S_{22}-S_{2-2}), \ \ \, \Gamma_5= S_{22}+S_{2-2}\nn \\
\Gamma_2&=&-S_{21}+S_{2-1}, \ \ \, \Gamma_3= i(S_{21}+S_{2-1}) \nn \\
\Gamma_4&=& \sqrt 2 S_{20},
\eea
which satisfy the anti-commutation relation 
\bea
\Gamma^a\Gamma^b+\Gamma^b\Gamma^a=2\delta_{ab}.
\eea

The rank-3 spin tensors $S_{3,m}$, also called spin-octupole 
operators, are constructed as follows
\begin{widetext}
\bea
S_{33}&=&\frac{\sqrt 2}{3} S_{11}^3=-\left(\begin{array}{cccc}
0&0&0&1\\
0&0&0&0\\
0&0&0&0\\
0&0&0&0
\end{array}\right),  
\hspace{25mm}
S_{32}=\frac{1}{\sqrt{6}} [S_-, S_{33}]=
\frac{\sqrt{2}}{2}\left(\begin{array}{cccc}
0&0&1&0\\
0&0&0&-1\\
0&0&0&0\\
0&0&0&0
\end{array}\right), \nn \\
S_{31}&=& \frac{1}{\sqrt{10}} [S_-, S_{32}]=
-\frac{1}{\sqrt{5}}\left(\begin{array}{cccc}
0&1&0&0\\
0&0&-\sqrt{3}&0\\
0&0&0&1\\
0&0&0&0
\end{array}\right),  \ \ \,
S_{30}= \frac{1}{2 \sqrt{3}} [S_-, S_{31}]=
\frac{1}{2 \sqrt{5}}\left(\begin{array}{cccc}
1&0&0&0\\
0&-3&0&0\\
0&0&3&0\\
0&0&0&-1
\end{array}\right),\nn \\
S_{3,-1}&=& \frac{1}{2 \sqrt{3}}[S_-,S_{30}]= -S_{31}^\dagger, 
\hspace{32mm}
S_{3,-2}=\frac{1}{\sqrt{10}} [S_-,S_{3,-1}]=S_{32}, \nn \\
S_{3,-3}&=&\frac{1}{\sqrt{6}}[S_-,S_{3,-2}]=-S_{33}^\dagger
\eea
\end{widetext}

\section{Pairing matrices for the isotropic $p$, $d$, and $f$-wave}
\label{sect:matrix}

By using the spherical harmonic functions $Y_{lm}(\hat k)$ and spin-tensors, 
we can construct the pairing matrices for the isotropic pairings for 
the spin-$\frac{3}{2}$ fermions in the $p$, $d$, and $f$-wave channels, 
respectively. 
The pairing matrix $\Delta_{\alpha\beta}(\vec k)$ in momentum space
can be represented as
\bea
\Delta^l_{\alpha\beta}(\vec k)&=& c_{l} 
\Delta
\left(\frac{k}{k_f} \right)^l(-)^m Y_{lm}(\hat k) S^{lm}R 
\nn \\
&=& c_{l} \Delta \left(\frac{k}{k_f}\right)^l  Y_{lm}^* (\hat k) S^{lm} R,
\eea
in which $l=1,2,3$ represent $p$, $d$, and $f$-wave pairings,
respectively, while $c_{l}$ is an overall constant factor.
These pairing structures are isotropic in analogy to the
$^3$He-B phase: the pairing orbital angular momenta are $l$,
and the pairing spins are also $l$, such that they
add together into the channel of total angular momentum 
$J=0$. 
The matrix kernel $Y_{lm}^*(\hat k) S^{lm}$ can be explicitly represented 
in isotropic forms as
\bea
Y_{lm}^*(\hat k) S^{lm} =\left\{
\begin{array}{ll}
\vec k \cdot \vec S , & (l=1)\\
\frac{1}{ \sqrt{2}}  (\vec k \cdot \vec S)^2 -\frac{5}{4 \sqrt{2}} k^2, & (l=2) \\
\frac{\sqrt{5}}{3} (\vec k \cdot \vec S)^3 -\frac{41}{12\sqrt{5}}k^2 (\vec k \cdot \vec S),
&(l=3)
\end{array}
\right. \nn \\
\eea

More explicitly, the pairing matrix in momentum space can be expressed
as follows.
In the $p$-wave case ($c_{1}=1$), it is
\bea
\Delta^p_{\alpha\beta}(\vec k)=\frac{\Delta}{k_f}
\left(\begin{array}{cccc}
0&0&-\frac{\sqrt 3}{2}k_-&\frac{3}{2}k_z\\
0&k_-&-\frac{1}{2}k_z&\frac{\sqrt 3}{2}k_+\\
-\frac{\sqrt 3}{2}k_-&-\frac{1}{2}k_z&-k_+&0\\
\frac{3}{2}k_z &\frac{\sqrt 3}{2}k_+&0&0
\end{array}
\right). \nn \\
\eea
\begin{widetext}
In the $d$-wave case ($c_{2}=2 \sqrt{2}$), it reads
\bea
\Delta^d_{\alpha\beta}(\vec k)=\frac{\Delta}{k_f^2}
\left(\begin{array}{cccc}
0 & \sqrt{3} k_{-}^{2} & -2\sqrt{3} k_{-}k_{z} & 3 k_{z}^{2}-k^{2} \\
-\sqrt{3} k_{-}^{2} & 0 & 3k_{z}^{2}-k^{2} & 2 \sqrt{3} k_{+}k_{z}\\
2 \sqrt{3} k_{-}k_{z} & k^{2}-3k_{z}^{2} & 0 & \sqrt{3} k_{+}^{2}\\
k^{2}-3k_{z}^{2} & -2\sqrt{3} k_{+} k_{z} & -\sqrt{3} k_{+}^{2} & 0
\end{array}
\right),
\eea
and in the $f$-wave case ($c_{3}= -\frac{3\sqrt{5}}{2}$), it becomes
\bea
\Delta_{\alpha\beta}(\vec k)=\frac{3\sqrt{3}\Delta }{8 k_f^3}  
\left(\begin{array}{cccc}
\frac{5}{\sqrt{3}}k_{-}^{3} & -5k_{-}^{2}k_{z} & 
-k_{-}(k^{2}-5k_{z}^{2}) & \frac{1}{\sqrt{3}} (3k^{2}-5k_{z}^{2})k_{z}\\
-5k_{-}^{2}k_{z} & -\sqrt{3}k_{-}(k^{2}-5k_{z}^{2}) & 
\sqrt{3} (3k^{2}-5k_{z}^{2})k_{z} & k_{+} (k^{2}-5k_{z}^{2})\\
-k_{-}(k^{2}-5k_{z}^{2}) & \sqrt{3} (3k^{2}-5k_{z}^{2})k_{z} 
& \sqrt{3} k_{+} (k^{2}-5k_{z}^{2}) & -5k_{+}^{2}k_{z}\\
\frac{1}{\sqrt{3}} (3k^{2}-5k_{z}^{2})k_{z} & k_{+} (k^{2}-5k_{z}^{2}) 
& -5k_{+}^{2}k_{z} & -\frac{5}{\sqrt{3}}k_{+}^{3}
\end{array}\right).
\eea

The general mean-field Hamiltonian in coordinate space is represented as
\bea
H=\frac{1}{2} \int  d^3 \vec r
\Psi^{\dagger}(\vec r)
& \left[ \begin{array}{cc}
(-\frac{\hbar^{2}}{2m} \nabla^{2}-\mu)I_{4} 
& c_{l} \frac{\Delta}{k_{f}^l} Y^*_{lm}(-i\nabla) S^{lm} R\\
c_{l} \frac{\Delta}{k_{f}^l}(Y^*_{lm}(-i\nabla) S^{lm} R )^{\dagger} 
& -(-\frac{\hbar^{2}}{2m}\nabla^{2}-\mu)I_{4}
\end{array}\right]
\Psi(\vec r),
\label{eq:pair}
\eea
\end{widetext}
in which $\Psi^\dagger(\vec r)=(c^\dagger_\alpha(\vec r), c_\beta(\vec r))$
is the Nambu spinor; $-i\nabla$ replaces $\vec k$ in the 
expressions of $k^lY_{lm}(\vec k)$.

\section{The symmetry properties}
\label{sect:symm}

Here we use the isotropic $f$-wave pairing state as an example
to illustrate the symmetry properties of this class of pairings.
The Hamiltonian Eq. 1 (main text)
processes both rotation
symmetries in the orbital and spin channels, {\it i.e.}, 
$SO_L(3)\times SO_S(3)$, while in the isotropic pairing state 
characterized by Eq. 5 (main text)
 only the total angular
momentum is conserved, {\it i.e.}, the residue symmetry is $SO_J(3)$.
The general configuration of the $d$-tensor can be expressed as
\bea
d_R^\nu(\hat k )=d^\nu(R^{-1} \hat k)=D^{l=3}_{\nu\nu^\prime}(R)
d^{\nu^\prime}( \hat k), 
\eea
where $R$ is an arbitrary SO(3) rotation and $D^l_{\nu\nu\prime}(R)$
is the rotation $D$-matrix. 
The relative SO symmetry is spontaneously broken similar
to the case of $^3$He-B, and here it is realized in a high
representation of angular momentum.
Combining with the $U(1)$ gauge symmetry breaking in the paired 
superfluid state, the Goldstone manifold is
$[SO_L(3)\otimes SO_S(3)\otimes U(1)]/SO_J(3)=SO(3)\otimes U(1)$.
Accordingly there exist four branches of Goldstone modes,
including one branch of phonon mode and three branches
of relative spin-orbit modes.

\section{Calculation of the bulk topological index}
\label{sect:nw}

In this section, we calculate the topological index of various
pairing states.
According to the definition of $Q(\vec k)=U^\dagger(\hat k) \Lambda(k)
U(\hat k)$ in which $U$ and $U^\dagger$ only depend on the direction
of $\vec k$, while $\Lambda(k)$ only depends on the magnitude of
$k$, we have
\bea
\nabla_k Q&=& U^\dagger \nabla_k \Lambda(k) U, \nn\\
\nabla_\theta Q&=& \nabla_\theta U^\dagger \Lambda U +U^\dagger \Lambda
\nabla_\theta U, \nn \\
\nabla_\phi Q&=& \nabla_\phi U^\dagger \Lambda U +U^\dagger \Lambda \nabla_\phi U,
\eea
in which $\nabla_k=\hat k \cdot \nabla$, 
$\nabla_\theta=\hat e_{\theta_k} \cdot \nabla$,
$\nabla_\phi=\hat e_{\phi_k} \cdot \nabla$.
Substituting the above equations into Eq. 8 in the main text,
after simplification, we arrive at
\bea
N_w&=&\frac{1}{4\pi^2}\int d^3\vec k \mbox{Tr}
(\nabla_\theta U \nabla_\phi U^\dagger -\nabla_\phi U \nabla_\theta U^\dagger)
\nabla_k \Lambda \Lambda^\dagger, \nn \\
&=&\sum_\lambda q_\lambda w_\lambda,
\eea
in which $q_\lambda$ is the monopole charge associated to the 
Berry curvature of the helicity eigenstate;
the corresponding eigenvalue $\lambda$ is
defined as
\bea
q_\lambda&=& \int \frac{k^2 d\Omega_k}{4\pi} F^{\lambda}_{\theta\phi}(\hat k)\nn\\
&=&\int \frac{k^2 d\Omega_k}{4\pi} (-i) \left(\nabla_\theta 
U \nabla_\phi U^\dagger-
\nabla_\phi U \nabla_\theta U^\dagger\right)_{\lambda\lambda} \nn \\
&=&\lambda,
\eea
and $w_\lambda$ is the winding number of the angular $\theta_\lambda(k)$
along the radial direction of $k$ as
\bea
w_\lambda=\int_0^{+\infty} \frac{dk}{\pi} 
~\left( i \nabla \Lambda \Lambda^\dagger \right)_{\lambda\lambda}
= \left\{ \begin{array}{l}
\mbox{sgn}(\xi_\lambda) ~~(\mu>0), \\
0 ~~(\mu<0).\\
\end{array}
\right. \nn \\
\eea
Consequently, we arrive at
\bea
N_w =\left \{ \begin{array}{l}
\sum_\lambda \lambda ~\mbox{sgn}(\xi_\lambda)  ~~(\mu>0), \\
0 ~~(\mu<0).
\end{array}
\right. 
\eea

\section{The surface modes of the $f$-wave isotropic pairing}
\label{sect:f_surface}

In this part, we study the gapless surface states of the
$f$-wave septet pairing.

We study a boundary imposed at $z=0$ with a spatial dependent 
chemical potential $\mu(z)$: $\mu_L=\frac{\hbar k_f^2}{2m}>0$ 
at $z<0$ and $\mu_R<0$ at $z>0$. 
For simplicity, we consider the case of $|\mu_R|\gg \mu_L$ and finally take
the limit of $|\mu_R|\rightarrow \infty$, i.e., at $z>0$ is
the vacuum. 

\subsection{The $f$-wave zero modes at $\vec k_\pp=0$}
To warm up, we first consider the case of $\vec k_\parallel=0$ in which
$S_z$ remains a good quantum number, and the zero modes described by 
different $S_z$ eigenvalues decouple.
The B-deG equation of the zero mode with $S_z$-eigenvalue 
$\alpha$ becomes
\bea
\left( \begin{array}{cc}
-\frac{\hbar^2}{2m}\frac{d^2}{dz^2}-\mu(z) & -i\frac{\Delta_{\alpha}}{k_f^3}
\frac{d^3}{dz^3}\\
-i\frac{\Delta_{\alpha}}{k_f^3} \frac{d^3}{dz^3} & 
\frac{\hbar^2}{2m}\frac{d^2}{dz^2}+\mu(z)
\end{array}
\right)
\left(\begin{array}{c}
u^0_\alpha(z)\\
v^0_{-\alpha}(z)
\end{array}
\right)=0,
\nn \\
\eea
in which $\Delta_{\pm\frac{3}{2}}=\frac{3}{4} \Delta$ and 
$\Delta_{\pm\frac{1}{2}}=\frac{9}{4} \Delta$.
The boundary condition is that
\bea
u_0(z)\rightarrow 0, \ \ \, v_0(z)\rightarrow 0, 
\label{eq:boundary}
\eea
as $z\rightarrow \pm \infty$. 

Eq. 10 (main text)
is invariant under the operation of 
$\left(\begin{array}{c} u_0 \\ v_0\end{array}\right)\rightarrow
i\tau_2 \left(\begin{array}{c} u_0 \\ v_0\end{array}\right)$ 
in which $\tau_2$ acts in the Nambu space, thus we can 
set $v_0=\pm i u_0$.
As it will be clear later that the solution actually satisfies 
$v_0(z)=-iu_0(z)$, 
the other one with $v_0(z)=i u_0(z)$ corresponds to the case
that the system lies at $z>0$ and the vacuum is at $z<0$. 
Then the equation becomes
\bea
\left(-\frac{\hbar^2}{2m}\frac{d^2}{dz^2}-\mu(z)-\frac{\Delta_\alpha}{k_f^3}
\frac{d^3}{dz^3}\right)u_0(z)=0.
\label{eq:reduced boundary}
\eea
We try the solution $u_0(z)\sim e^{\beta_L z}$ at $z<0$, and, $e^{\beta_R z}$
at $z>0$, and thus $\Re \beta_L >0$ and $\Re\beta_R<0$, respectively.

At $z<0$, $\beta_L$ satisfies the cubic equation with real coefficients
\bea
\left(\frac{\beta_L}{k_f}\right)^3+\frac{\epsilon_f}{\Delta_\alpha} 
\left(\frac{\beta_L}{k_f}\right)^2 +\frac{\epsilon_f}{\Delta_\alpha}=0,
\eea
which has a pair of conjugate complex roots and one real root. 
We only consider the weak pairing limit that $\frac{\Delta}{\epsilon_f}\ll 1$.
The solutions correct to the linear order of $\frac{\Delta}{\epsilon_f}\ll 1$ 
are 
\bea
\left(\frac{\beta_L}{k_f}\right)_{1,2}\approx\pm i +\frac{\Delta_\alpha}{2\epsilon_f}, \ \ \,
\left(\frac{\beta_L}{k_f}\right)_{3}\approx -\frac{\epsilon_f}{\Delta_\alpha},
\eea
and thus only $(\frac{\beta_L}{k_f})_{1,2}$ can be kept. 
Similarly, at $z>0$, in the case of $|\mu_R|\gg \epsilon_f$, there exists
a pair of complex conjugate roots and one real root for $\beta_R$ as
\bea
\left(\frac{\beta_R}{k_f}\right)_{1,2}\approx c
\left(-\frac{1}{2}\pm i \frac{\sqrt 3}{2}\right), \ \ \, (\frac{\beta_{R}}{k_{f}})_{3} \approx c,
\eea
in which $c=\left(|\mu_R|/ \Delta_{\alpha} \right)^{\frac{1}{3}}$.

Because 
Eq. 10 (main text)
is a 3rd order differential equation, 
all of $u_0(z), \frac{d}{dz}u_0(z), \frac{d^2}{dz^2} u_0(z)$ need
to be continuous at the boundary $z=0$.
For this purpose, we construct the following solution
\bea
u_0(z)=\left\{\begin{array}{cc}
A_L \sin(k_f z+\phi_L) e^{\frac{\Delta_\alpha}{2\epsilon_f}k_f z} & (z<0)\\
A_R \sin(\frac{\sqrt 3}{2} c k_f z+\phi_R) e^{-\frac{c}{2} k_f z} & (z>0)
\end{array}
\right., \ \ \,
\label{eq:match}
\eea
in which the four parameters $A_{L(R)}$ and $\phi_{L(R)}$ are
sufficient to match three continuous conditions.
In the case of $c\rightarrow +\infty$, the results can be simplified as
\bea
\phi_L&=&0, \ \ \,
\phi_R=-\frac{\pi}{3} \nn \\
\frac{A_R}{A_L}&=& \frac{1}{c\sin(\frac{\pi}{3}-\phi_R)}
\rightarrow 0,
\eea
which shows that we can simply set $u_L(z)$ vanishing at
$z=0$. 

To summarize, we have solved 
\bea
u_{f,\alpha}(z)= \frac{1}{\sqrt{N_{\alpha}}} e^{\beta_\alpha z}\sin k_fz
\eea
with $\beta_{\pm \frac{3}{2}} = \frac{1}{3}\beta_{\pm \frac{1}{2}}
=\frac{3}{8} \frac{|\Delta|}{\epsilon_f} k_f$, and $N_{\pm \frac{3}{2}}
=3N_{\pm_{\frac{1}{2}}} = \frac{4}{3} \frac{\epsilon_f}{ |\Delta|} \frac{1}{k_{f}}$.

\subsection{The $k\cdot p$ perturbation theory for midgap states}

The effective Hamiltonian for the midgap states on the surface of
the $f$-wave isotropic pairing is presented in Eq. 11 (main text).
The spectra consist of two gapless Dirac cones denoted as $a$ and $b$,
respectively, as shown in the main text. 
The corresponding eigenfunctions are solved as
\bea
\psi^{a}_{\pm}(\vec k_\parallel)&=&
\frac{1}{\sqrt N}
\left( \begin{array}{c}
\mp i e^{-i \frac{3}{2}\phi_{k}}\\
-x e^{-i \frac{\phi_k}{2}  } \\
\pm i x e^{i \frac{\phi_k}{2} }\\ 
e^{i \frac{3}{2} \phi_{k} },
\end{array}
\right), \nn \\
\psi^{b}_{\pm}(\vec k_\parallel)&=&\frac{1}{\sqrt N}
\left(\begin{array}{c}
\pm i x e^{-i \frac{3}{2} \phi_{k} }, \\ 
e^{-i \frac{1}{2} \phi_{k} } \\
\pm i e^{i \frac{1}{2} \phi_{k} } \\
x e^{i \frac{3}{2} \phi_{k} }, 
\end{array}
\right),
\eea
in which $x=\sqrt 2+1$; $N=2\sqrt{2+\sqrt 2}$; $\phi_k$ 
is the azimuthal angle of $\vec k_\parallel$.

The eigen-solutions $\psi^{a(b)}_\pm(\vec k_\pp)$ are parity eigenstates 
of the little group symmetry of the reflection $\sigma_v(\vec k_\pp)$,
which is defined with respect to the vertical plane passing $\vec k_\pp$
and the $z$-axis $\hat z$.
The operation $\sigma_v(\vec k_\pp)$ can be decoupled as
a combined operation of inversion and rotation as
\bea
\sigma_v(\vec k_\pp)=i I R_{\vec k^\prime}(\pi)
= \left(\begin{array}{cccc}
0 &0 &0 & -i e^{-i3\phi_{k}} \\
0 &0 &ie^{-i\phi_{k}} &0 \\
0 &-e^{i\phi_{k}} &0 &0 \\
ie^{3\phi_{k}} &0 &0 &0
\end{array} 
\right),\nn \\
\label{eq:parity}
\eea
in which $I$ is the inversion operation; $\phi_{k}$ is the azimuthal angle of $\vec{k}_{\parallel}$; $\vec k^\prime$ is
an in-plane momentum perpendicular to $\vec k_\pp$ and 
$R_{\vec k^\prime}(\pi)$ is rotation around $\vec k_\pp$ at the
angle of $\pi$; the factor of $i$ is to make $\sigma_v$ an
Hermitian operator with eigenvalues $\pm 1$. 
It is easy to check that 
\bea
\sigma_v (\vec k_\pp) \psi^a_\pm (\vec k_\pp) &=& \pm  \psi^a_\pm (\vec k_\pp),
\nn \\
\sigma_v(\vec k_\pp) \psi^b_\pm (\vec k_\pp) &=& \mp  \psi^b_\pm (\vec k_\pp),
\eea
respectively.

\subsection{The surface zero energy ring states}

Here we present the eigenfunctions of the zero energy ring of the 
midgap surface states of the isotropic $f$-wave pairing state, 
which is located at $k_\pp^0=\frac{\sqrt 3}{2}k_f$. 
The method of solution can be referred to SM
\ref{sect:general}.
The zero energy states at $\vec k_\pp= k_\pp^0 (\cos\phi_k,\sin\phi_k)$
are two-fold degenerate, whose creation operators are denoted as 
$\gamma_{1,2}(\vec k_\pp)$, respectively.
They are explicitly expressed below as
\begin{widetext}
\bea
\gamma_{1}^\dagger(\vec k_\pp)=\int^0_{-\infty}dz ~ 
\sum_{\alpha}
\left[e^{i(\frac{\phi}{2}+\frac{\pi}{4})}c^\dagger_\alpha(\vec k_\pp,z)
+ (-)^{\alpha+\frac{1}{2}} 
e^{-i(\frac{\phi}{2}+\frac{\pi}{4})}c_\alpha(-\vec k_\pp,z)\right]
e^{-i\alpha\phi_k} u_{\alpha}(z), \nn \\
\gamma_{2}^\dagger(\vec k_\pp)=\int^0_{-\infty}dz ~ 
\sum_{\alpha}
\left[ (-)^{\alpha-\frac{1}{2}} 
e^{i(\frac{\phi}{2}+\frac{\pi}{4})}c^\dagger_\alpha(\vec k_\pp,z)
+ 
e^{-i(\frac{\phi}{2}+\frac{\pi}{4})}c_\alpha(-\vec k_\pp,z)\right]e^{-i\alpha\phi_k}
u_{-\alpha}(z),
\eea
\end{widetext}
in which $\phi$ is the phase of the pairing amplitude $\Delta$;
$\alpha=\pm\frac{3}{2},\pm\frac{1}{2}$ as the eigenvalue of
$S_z$; the envelope  wavefunctions $u_{\alpha}(z)$ are
\bea
u_{\frac{3}{2}}(z)&=& \frac{1}{\sqrt{N_{\frac{3}{2}}}} 
3\cos(\frac{k_fz}{2}) (e^{\beta_{1} z}-e^{\beta_{2} z}),\nn \\
u_{\frac{1}{2}}(z)&=& \frac{1}{\sqrt{N_{\frac{1}{2}}}} \sin(\frac{k_f z}{2}) (3 e^{\beta_{1}z} + 5 e^{\beta_{2} z}) ,\nn \\
u_{-\frac{1}{2}}(z)&=& \frac{1}{\sqrt{N_{-\frac{1}{2}}}} \sqrt{3} \cos(\frac{k_fz}{2}) (e^{\beta_{1} z} - e^{\beta_{2} z}) , \nn \\
u_{-\frac{3}{2}}(z)&=& \frac{1}{\sqrt{N_{-\frac{3}{2}}}} 
\sin(\frac{k_f z}{2}) ( \frac{1}{\sqrt{3}} e^{\beta_{1}z} 
-3 \sqrt{3} e^{\beta_{2} z} ),  
\nn \\
\eea
in which $\beta_2=3\beta_1=\frac{9}{4}\frac{|\Delta|}{k_f}$,
and $N_{\alpha}$'s are the overall normalization factors whose
expressions are complicated and will not be presented.

Since $\gamma^\dagger_{1,2}(\vec k_\pp)$ represent the zero energy
modes, again they are chiral eigen-modes satisfying
\bea
C_{ch} \gamma_{1}(\vec k_\pp) C_{ch}^{-1} &=&\gamma_{1}(\vec k_\pp), \nn \\
C_{ch} \gamma_{2}(\vec k_\pp) C_{ch}^{-1} &=&-\gamma_{2}(\vec k_\pp). 
\eea
Nevertheless, they are not parity eigen-modes, and
they transform into each other under the parity operation defined 
in Eq. \ref{eq:parity} as
\bea
\sigma_{v}(\vec k_\pp)\gamma_1(\vec k_\pp )\sigma_{v}^{-1}(\vec k_\pp)
&=& \gamma_2(\vec k_\pp ).
\eea

\section{The surface states of the $p$-wave isotropic pairing}
\label{sect:p_surface}

In this section, we consider the surface states of the $p$-wave
case under the same planar boundary configuration as that in the 
$f$-wave case.

\subsection{The $p$-wave zero modes at $\vec k_\pp=0$}

There also exists one zero mode at $\vec k_\pp=0$ for each spin 
component in the $p$-wave case, whose spatial wavefunctions 
will be solved below.

The equation parallel to Eq. \ref{eq:reduced boundary} is 
\bea
\left(-\frac{\hbar^2}{2m}\frac{d^2}{dz^2}-\mu(z)
+ \frac{|\Delta_{\alpha}|}{k_{f}} \frac{d}{dz} \right)u_0(z)=0.
\label{eq:p boundary}
\eea
in which $\Delta_{\pm \frac{3}{2}}=\frac{3}{2} \Delta$, 
$\Delta_{\pm \frac{1}{2}}=-\frac{1}{2} \Delta$, and 
$v_{0}(z)=i u_{0} (z)$ for $\alpha=\pm \frac{3}{2}$, $v_{0}(z)=-iu_{0}(z)$ 
for $\alpha=\pm \frac{1}{2}$. 
Again we set $u_0(z)\sim e^{\beta_L z}$ at $z<0$, and, $e^{\beta_R z}$ at $z>0$, 
with $\Re \beta_L >0$ and $\Re\beta_R<0$.
At $z<0$, $\beta_{L}$ satisfies the equation
\bea
\left(\frac{\beta_{L}}{k_{f}}\right)^{2} - 
\frac{|\Delta_{\alpha}|}{\epsilon_{f}} \left(\frac{\beta_{L}}{k_{f}}\right) +1=0,
\eea
which in the limit $\frac{\Delta}{\epsilon_{f}}<<1$ has the solutions 
\bea
\left(\frac{\beta_{L}}{k_{f}}\right)_{1,2} 
\approx \pm i + \frac{|\Delta_{\alpha}|}{2\epsilon_{f}}.
\eea
At $z>0$, in the limit of $|\mu_{R}|>>\epsilon_{f}$, $\beta_{R}$ 
has two solutions as
\bea
\beta_{R,1,2} \approx \pm \left(\frac{|\mu_{R}|}{\epsilon_{f}}\right)^{\frac{1}{2}},
\eea
and the negative one is kept to match boundary condition at 
$z \rightarrow +\infty$.

Eq. \ref{eq:p boundary} is a second order differential equation, 
and thus $u_{0} (z)$ and $\frac{d}{dz}u_{0} (z)$ need to be 
continuous at $z=0$. 
Similar to the $f$-wave case, we arrive at
\bea
u_{p,\alpha} (z) = \frac{1}{\sqrt{N_{\alpha}}} e^{\beta_{\alpha} z} \sin k_{f} z,
\eea
with $\beta_{\alpha} = \frac{|\Delta_{\alpha}|}{2 \epsilon_{f}} k_{f}$,
and $N_{\alpha} = \frac{\epsilon_{f}}{|\Delta_{\alpha}|} \frac{1}{k_{f}}$.

\subsection{The $k\cdot p$ perturbation theory}

The chiral indices for the zero modes $|0_\alpha\rangle_p$ at $\vec k_\pp=0$
are $1, 1, -1, -1$ for $\alpha=\frac{3}{2},\frac{1}{2},-\frac{1}{2},
-\frac{3}{2}$, respectively.
We can use these zero modes as the bases to construct the effective
Hamiltonian for low energy midgap states at $\vec k_\pp\ll k_f$
with respect to $\delta H=H(\vec k_\pp, z) -H(0,z)$.
As constrained by the surface symmetry $C_{v\infty}$ and the chiral 
symmetry of the zero modes, the effective Hamiltonian is
\bea
H_{mid}^p(\vec k_\parallel) = \frac{\Delta }{k_f} \left(
\begin{array}{cccc}
0& 0 & c k_-^2 & O (k_-^3 )\\
0& 0 & -i k_- & c k_-^2 \\
c k_+^2&  i k_+ &0 & 0 \\
O(k_+^3)&c k_+^2 & 0 &0
\end{array}
\right), \ \ \,
\label{eq:pmidgap}
\eea
in which the terms proportional to $k^3_\pm$ arise from the 
3rd order perturbation theory and are neglected. 
The terms proportional to $k^2_\pm$ are due to the 2nd
order perturbation theory involving gapped bulk states
as intermediate states.
Under a suitable phase convention $c$ is a real coefficient
at the order of $1/k_f$, whose concrete value is not important.
At the leading order, the two components with $\alpha=\pm\frac{1}{2}$ form 
a linear Dirac cone, while the other two with $\alpha=\pm\frac{3}{2}$
are dispersionless. 
Nevertheless, the latter are coupled indirectly 
through coupling with the former and develop a cubic Dirac cone.

\section{The isotropic $d$-wave pairing}
\label{sect:d_surface}

The isotropic $d$-wave pairing with spin-$\frac{3}{2}$ fermions is 
actually topologically trivial. 
In this part, we explicitly check this point from the boundary spectra. 
The boundary configuration is the same as before, and
we will show the absence of the zero modes at $\vec k_\pp=0$. 

Similar to the $f$- and $p$-wave cases, the equation determining zero 
modes is invariant under $i\tau_{2}$ operation. 
Let $v_{0}(z)=i\eta u_{0}(z)$ ($\eta=\pm 1$), we obtain 
\bea
\left(- \frac{d^{2}}{dz^{2}} - \frac{\mu(z)}{\epsilon_{f}} k_{f}^{2}
\right) u_{0}(z) 
- i \eta \frac{\Delta_{\alpha}}{\epsilon_{f}} \frac{d^{2}}{dz^{2}} u_{0} (z) = 0,
\label{eq:d boundary}
\eea
in which $\Delta_{\frac{3}{2}}=\Delta_{\frac{1}{2}} = 2 \Delta$, 
$\Delta_{-\frac{3}{2}}=\Delta_{-\frac{1}{2}} = - 2 \Delta$.
Expressing $u_0(z)\sim e^{\beta_L z}$ at $z<0$, and, $e^{\beta_R z}$ at $z>0$, 
$\beta_{L}$ and $\beta_{R}$ are solved as
\bea
\frac{\beta_{L}}{k_{f}} &\approx& \pm \left (1- \frac{1}{2} i 
\eta \frac{\Delta_{\alpha}}{\epsilon_{f}}\right), \nn \\
\frac{\beta_{R}}{k_{f}} 
&\approx& \pm \sqrt{\frac{|\mu_{R}|}{\epsilon_{f}}} 
\left (1- \frac{1}{2} i \eta \frac{\Delta_{\alpha}}{\epsilon_{f}}\right),
\eea
where $|\Delta_{\alpha}| \ll \epsilon_{f} \ll |\mu_{R}|$ is assumed. 

Since Eq. \ref{eq:d boundary} is a 2nd order differential equation, 
both $u_{0} (z)$ and $\frac{d}{dz} u_{0} (z)$ need to be continuous at $z=0$. 
Regardless of value of $\eta$, there is only one $\beta_{L}$ with a positive 
real part, and one $\beta_{R}$ with a negative real part. 
The boundary conditions at $z=0$ are two linear homogeneous equations 
of the undetermined coefficients of wavefunctions.
Generally speaking, there is only zero solution, which
demonstrates the absence of zero modes for $d$-wave isotropic
pairing.

\section{The method for solving the midgap surface states}
\label{sect:general}

In this section, we present a general method for solving surface 
states in the weak pairing limit away from the $\vec{k}_{\parallel}=0$ point. 
The isotropic $f$-wave pairing in the spin-$\frac{3}{2}$ system
is used as an example, and actually the method can be directly 
applied to other partial-wave channels and higher spins.  

\subsection{Match boundary conditions}
Consider the same boundary configuration as stated before in 
Supp. Mat. \ref{sect:f_surface}.
We denote $\Phi(\vec{r})$ the eigen-wavefunction with the eigen-energy
$E$ and the in-plane wavevector $\vec k_\pp=(k_{x},k_{y})$.
The following trial solution will be used
\bea
\Phi(\vec{r}) \sim 
\left\{
\begin{array}{l}
 \Phi^{L} e^{\beta^{L} z} e^{i k_{x} x +i k_{y} y}, \ \ \, ~~ z<0,  \\
\Phi^{R} e^{\beta^{R} z} e^{i k_{x} x +i k_{y} y}, \ \ \, z>0,
\end{array}\right.
\label{eq:wavefunction}
\eea
in which $\Phi^{L}$, $\Phi^{R}$ are 8-component column vectors.
Denote $H_{L}$ and $H_R$ the Hamiltonians for $z<0$ and $z>0$
with the corresponding chemical potentials $\mu_{L}$ and $\mu_R$, respectively.
Substituting the trial wavefunction 
into the eigen-equations at $z<0$ and $z>0$, 
respectively, the conditions for the existence of nonzero
solutions of $\Phi^{L}$ and $\Phi^{R}$ are obtained as
\bea
\left\{
\begin{array}{l}
\det [H_{L}(k_{x},k_{y},-i\beta^{L})-E] = 0, \\
\det [H_{R}(k_{x},k_{y},-i\beta^{R})-E] = 0.
\end{array}
\right.
\label{eq:determinant}
\eea

Both solutions $-i\beta^{L}$ and $-i\beta^{R}$ appear in terms of
complex conjugate pairs, since the determinant 
Eq. \ref{eq:determinant} are real equations.
Consequently, among the $24$ solutions of $\beta_j^{L}$, there are $12$ 
solutions with positive real parts and $12$ with negative real parts, 
and so do $\beta_j^{R}$'s.
The midgap state needs to vanish at $z\rightarrow \pm \infty$, 
hence, it is in the form of
\bea
\Phi(\vec{r})=
\left\{
\begin{array}{l}
\sum_{j=1}^{12} B^{L}_{j} \Phi^{L}_{j} e^{\beta^{L}_{j} z} e^{i(k_{x}x+k_{y}y)}, \ \ \, z<0, \\
\sum_{j=1}^{12} B^{R}_{j} \Phi^{R}_{j} e^{\beta^{R}_{j} z} e^{i(k_{x}x+k_{y}y)}, \ \ \,z>0,
\end{array}
\right. 
\label{eq:wavef}
\eea
in which $\Re \beta^{L}_{j}>0$ and $\Re \beta^{R}_{j}<0$, $1\leq j \leq 12$.

The boundary conditions require the wavefunctions Eq. \ref{eq:wavef}, 
and their first and second order derivatives to be continuous at $z=0$.
We have a set of linear homogeneous equations that
the coefficients $B^{L}_{j}$ and $B^{R}_{j}$ should obey.
The conditions for the existence of nonzero solutions are
\bea
\text{det} \left( \begin{array}{cc}
E_{L} & E_{R} \\
F_{L} & F_{R} \\
G_{L} & G_{R}
\end{array} \right)= 0
\label{eq:24times24}
\eea
in which $E_{L,R}$, $F_{L,R}$ and $G_{L,R}$ are $8\times 12$ rectangular
matrices, and thus the total dimension is $24\times 24$. 
The above block rectangular matrices are expressed as
\bea
E_{L}(\cdot,j) &=& \Phi^{L}_{j}, \hspace {12mm} E_{R}(\cdot,j) = \Phi^{R}_{j},
\nn \\
F_{L}(\cdot,j) &=& \beta^{L}_{j} \Phi^{L}_{j}, \hspace{8mm}
F_{R}(\cdot,j) = \beta^{R}_{j} \Phi^{R}_{j},   \nn \\
G_{L}(\cdot,j) &=& (\beta^{L}_{j})^{2} \Phi^{L}_{j},  \ \ \, 
G_{R}(\cdot,j) = (\beta^{R}_{j})^{2} \Phi^{R}_{j},
\eea
in which $(\cdot,j)$ denotes $j$-th column of the corresponding 
matrix.
Surface energies can be solved from this equation. 

Actually the complicated determinant equation Eq. \ref{eq:24times24}
can be greatly simplified in the weak pairing limit
as will be shown in Sect. \ref{sect:midgap}.

\subsection{Equations of the midgap state energy}
\label{sect:midgap}

In the half space at $z<0$, we rewrite the eigen-solution $\Phi(\vec r)$
in Eq. \ref{eq:wavef} as
\bea
\Phi(\vec{r})=\sum_{j=1}^{12}B^{L}_{j} \Phi^{L}_{j} e^{ik_{z,j} z} e^{i(k_{x}x+k_{y}y)},
\eea
in which $k_{z,j}=-i\beta^{L}_{j}$ ($1\leq j \leq 12$), 
hence, $\Im k_{z,j}<0$ such that $\Phi(\vec r)$ vanishes at
$z\rightarrow -\infty$.

It can be shown that the twelve $k_{z,j}$'s can be 
classified into two groups.
In one group, their real parts are very close to 
$\pm \sqrt{k_{f}^{2}-k_{\parallel}^{2}}$ as
\bea
k_{z,m}^{\pm}&=&\pm \sqrt{k_{f}^{2}-k_{\parallel}^{2}}-i\xi^{\pm}_m 
\Delta, 
\eea
in which $m=\pm\frac{3}{2},\pm\frac{1}{2}$,
and $\Phi^{L \pm}_{m}$ are corresponding eigen-vectors.
The remaining four $k_{z}$'s represent fast decaying modes in the
weak pairing limit, which are proportional to 
$-i \frac{\epsilon_{f}}{\Delta} k_{f}$.
It can also be proved that at the leading order of
$\frac{\Delta}{\epsilon_{f}}$, 
the wavefunction at $z<0$ is represented as
\bea
\Phi(\vec{r})=\sum_{m} 
\Big\{
B^{L+}_{m} \Phi^{L+}_{m} e^{i k^{+}_{z,m} z} + B^{L-}_{m} \Phi^{L-}_{m} 
e^{i k^{-}_{z,m} z}
\Big\} \ \ \,
\label{eq:wavefunctionsimplified},
\eea
in which the fast decaying mode contributions are neglected. 

It can be shown that in the limits of $\Delta \ll \epsilon_{f} 
\ll |\mu_{R}|$, the boundary conditions can be further simplified as 
$\Phi(\vec r)=0$ at the boundary of $z=0$.
The detailed proof is rather complicated but straightforward,
and will be presented elsewhere.
This great simplification reduces the equation determining surface energies 
from the original $24\times24$ determinant condition of Eq. \ref{eq:24times24} 
to the following one of $8\times8$,
\bea
\det\left(\Psi^{+},\Psi^{-}\right)=0,
\label{eq:simplifiedboundary}
\eea
in which $\Psi^{\pm}$ are $8\times4$ matrices defined by 
$\Psi^{\pm}(\cdot,m+\frac{5}{2})=\Phi^{L\pm}_{m}$ 
($m=\pm\frac{3}{2},\pm\frac{1}{2}$).
Furthermore, due to the $SU(2)$ bulk rotation symmetry and the reflection 
symmetry $\sigma_{v}(\vec{k_{\parallel}})$ defined in Eq. \ref{eq:parity},
Eq. \ref{eq:simplifiedboundary} can be further simplified to two 
$4\times 4$ determinant equations. 

The surface midgap energies are smaller than $\Delta$, we express 
$E=\epsilon \Delta$.
To solve $k_{z,m}^{\pm}$ and $\Phi^{L\pm}_{m}$, 
$k^{\pm}_{z}=\pm \sqrt{k_{f}^{2}-k_{\parallel}^{2}}-i \xi^{\pm} \Delta$ 
is plugged  into the eigen-equation.
Keeping only the leading order of $\Delta$, we obtain 
\begin{widetext}
\begin{eqnarray}
& \left(\begin{array}{cc}
\mp 2i\Delta \sqrt{k_{f}^{2}-k_{\parallel}^{2}} \xi^{\pm} I_{4} 
& K(k_{x},k_{y},\pm  \sqrt{k_{f}^{2}-k_{\parallel}^{2}})R\\
(K(k_{x},k_{y},\pm  \sqrt{k_{f}^{2}-k_{\parallel}^{2}})R)^{\dagger} 
& \pm 2i\Delta \sqrt{k_{f}^{2}-k_{\parallel}^{2}} \xi^{\pm} I_{4}
\end{array}\right) 
\Phi^{L\pm} =\epsilon \Delta \Phi^{L\pm}.
\end{eqnarray}
\end{widetext}
in which the subscript $m$ is dropped for simplicity.
Since $K(k_{x},k_{y}, -i\partial_{z})$ already contains a prefactor
of $\Delta$, the $-i\partial_{z}$ can be substituted by
$\pm \sqrt{k_{f}^{2}-k_{\parallel}^{2}}$ without inducing
higher order error.
Denote $U_\pm(\vec k)$ as the rotation matrices associated with the operations
rotating $\pm \hat{z}$ to the direction of $\vec{k}$,
the pairing part can be represented as
\bea
K(\pm k_{f} \hat{z})R= U^{-1}_\pm (\vec k_\pm) ~ K(\vec k_\pm )R 
~U^{*}_\pm (\vec k_\pm),
\eea
in which $\vec k_\pm =(k_{x},k_{y},\pm \sqrt{k_{f}^{2}-k_{\parallel}^{2}})$.
Applying such rotation operations to the eigen-equation, we obtain
\begin{widetext}
\bea
\label{equation in z direction}
& \left(\begin{array}{cc}
\mp 2i \Delta \sqrt{k_{f}^{2}-k_{\parallel}^{2}} \xi^{\pm} I_{4} & K(\pm k_{f} \hat{z})R\\
(K(\pm k_{f} \hat{z})R)^{\dagger} & \pm 2i \Delta \sqrt{k_{f}^{2}-k_{\parallel}^{2}} \xi^{\pm} I_{4}
\end{array}\right)
\tilde{\Phi}^{L\pm} =\epsilon \Delta \tilde{\Phi}^{L\pm},
\eea
\end{widetext}
in which
\bea
\tilde{\Phi}^{L\pm}= W^{-1}_\pm(\vec k_\pm) \Phi^{L\pm}.
\eea
and $W$ is defined as
\bea
W_\pm(\vec k_\pm)=\left(\begin{array}{cc}
U_{\pm}(\vec k_\pm) & \\
& U^{*}_{\pm}(\vec k_\pm)
\end{array}\right).
\eea
Hence it is sufficient to solve $\tilde{\Phi}^{L\pm}$ to arrive at
$\Phi^{L\pm}$, which satisfy a simple equation where the pairing 
matrix is in $\pm \hat{z}$ direction. 

For notational convenience we define the column vectors
representing particle and hole states as
\bea
p_{\frac{3}{2}}&=&(1\,0\,0\,0\,,0\,0\,0\,0)^{T}, ~~
p_{\frac{1}{2}}=(0\,1\,0\,0\,0\,,0\,0\,0)^{T}, \nn \\
p_{-\frac{1}{2}}&=&(0\,0\,1\,0\,,0\,0\,0\,0)^{T},
p_{-\frac{3}{2}}=(0\,0\,0\,1\,,0\,0\,0\,0)^{T}, \nn 
\eea
and 
\bea
h_{\frac{3}{2}}&=&(0\,0\,0\,0\,,1\,0\,0\,0)^{T}, ~~ 
h_{\frac{1}{2}}=(0\,1\,0\,0\,, 0\,1\,0\,0)^{T}, \nn \\
h_{-\frac{1}{2}}&=&(0\,0\,1\,0\,, 0\,0\,1\,0)^{T},~~
h_{-\frac{3}{2}}=(0\,0\,0\,1\,, 0\,0\,0\,1)^{T}. \nn 
\eea
The solutions to $\xi^\pm_m$ and $\tilde{\Phi}_m^{L\pm}$ are 
summarized as follows (vectors un-normalized),
\begin{widetext}
\bea
\xi_{\frac{3}{2}}^{\pm}&=&\frac{\sqrt{\frac{9}{16} - \epsilon^{2} }}
{2 \sqrt{k_{f}^{2} - k_{\parallel}^{2} } } \Delta,  \ \ \, \ \ \,
\tilde{\Phi}_{\frac{3}{2}}^{\pm} = \frac{3}{4} p_{\frac{3}{2}} 
+ (-i \sqrt{\frac{9}{16} - \epsilon^{2} } \mp \epsilon ) h_{-\frac{3}{2}}, \nn \\
\xi_{\frac{1}{2}}^{\pm}&=&\frac{\sqrt{\frac{81}{16}- \epsilon^{2} } }{2 \sqrt{k_{f}^{2} - k_{\parallel}^{2} } } \Delta, \ \ \, \ \ \, 
 \tilde{\Phi}_{\frac{1}{2}}^{\pm} = \frac{9}{4}  p_{\frac{1}{2}} + (-i \sqrt{\frac{81}{16} - \epsilon^{2} } \mp \epsilon ) h_{-\frac{1}{2}}, \nn \\
\xi_{-\frac{1}{2}}^{\pm}&=& \xi_{-\frac{1}{2}}^{\pm}, 
\hspace{19mm}
\tilde{\Phi}_{-\frac{1}{2}}^{\pm} = \frac{9}{4}  p_{-\frac{1}{2}} + (-i \sqrt{\frac{81}{16} - \epsilon^{2} } \mp \epsilon ) h_{\frac{1}{2}}, \nn \\
\xi_{-\frac{3}{2}}^{\pm} &=&\xi_{\frac{3}{2}}^{\pm},
\hspace{21mm}
\tilde{\Phi}_{-\frac{3}{2}}^{\pm} = \frac{3}{4} p_{-\frac{3}{2}} + (-i \sqrt{\frac{9}{16} - \epsilon^{2} } \mp \epsilon ) h_{\frac{3}{2}}. 
\eea
\end{widetext}

Correspondingly, the determinant equation for the eigen-energies becomes
\bea
& \text{det}
\left[ W_+(\vec k_+)
\tilde{\Psi}^{L+},
W_-(\vec k_-)
\tilde{\Psi}^{L-} \right] = 0,
\eea
in which $\tilde{\Psi}^{L\pm}$ are $8\times4$ matrices defined 
by $\tilde{\Psi}^{\pm}(\cdot,m+\frac{5}{2}) = \tilde{\Phi}^{L\pm}_{m}$ 
($m=\pm\frac{3}{2},\pm\frac{1}{2}$).
As mentioned before, in this way, the original $24 \times 24$ 
matrix determinant equation is reduced to an $8 \times 8$ one. 

Further using the reflection symmetry, the above $8 \times 8$ matrix 
can be further decomposed into two $4 \times 4$ ones. 
Without loss of generality, we only consider $\vec k_\pp$ along the
$x$-axis, i.e., $\vec k_\pp =(k_\pp,0)$, and results for other 
values of $\vec k_\pp$ can be obtained by applying rotations around the $z$-axis. 
The reflection operator with respect to the vertical $xz$-plane 
which we denote by $\sigma_{vx}$, is given in the particle-hole 
$8$-dimensional space as
\bea
\sigma_{vx} = \left( \begin{array}{cc}
-iR & 0 \\
0 & iR
\end{array}\right).
\eea
The vectors $\tilde{\Phi}^{L\pm}_{m}$ can be recombined into even and odd
eigenvectors of $\sigma_{vx}$ defined as
\bea
\tilde{\Phi}^{\pm}_{e,\frac{3}{2}} &=&
\tilde{\Phi}^{L\pm}_{\frac{3}{2}} + i\tilde{\Phi}^{L\pm}_{-\frac{3}{2}}, \nn \\
\tilde{\Phi}^{\pm}_{e,\frac{1}{2}} &=& 
\tilde{\Phi}^{L\pm}_{\frac{1}{2}} - i\tilde{\Phi}^{L\pm}_{-\frac{1}{2}}, \nn \\
\tilde{\Phi}^{\pm}_{o,\frac{3}{2}} &=& 
\tilde{\Phi}^{L\pm}_{-\frac{3}{2}} - i\tilde{\Phi}^{L\pm}_{\frac{3}{2}}, \nn \\
\tilde{\Phi}^{\pm}_{o,\frac{1}{2}} &=&
 \tilde{\Phi}^{L\pm}_{\frac{1}{2}} + i\tilde{\Phi}^{L\pm}_{-\frac{1}{2}},
\eea
in which the subscripts ``$e$'' and ``$o$'' denote even and odd 
parity eigenvalues $1$ and $-1$ of $\sigma_{vx}$, respectively. 

For $\vec k_\pp=(k_\pp, 0)$, $U_{\pm}(\vec k_\pm)$ are rotations
around the $y$-axis, which commutes with $\sigma_{vx}$.
Applying a basis transformation $P$ which separates the 
even and odd parity eigen-spaces of $\sigma_{vx}$, we have
\bea
P^{-1} \tilde{\Phi}^{\pm}_{\text{e},\eta} &=& \left( \begin{array} {c}
\phi^{\pm}_{\text{e},\eta} \\
0
\end{array} \right),   \ \ \,
P^{-1} \tilde{\Phi}^{\pm}_{\text{o},\eta} = \left( \begin{array} {c}
0 \\
\phi^{\pm}_{\text{o},\eta}
\end{array} \right),    \ \ \,
\eea
where $\eta=\frac{3}{2},\frac{1}{2}$, and then
\bea
P^{-1} U_{\pm}(\vec k_\pm) P &=& \left( \begin{array}{cc}
U_{\pm,e}(\vec k_\pm) & 0 \\
0 & U_{\pm,o}(\vec k_\pm)
\end{array} \right). \ \ \, \ \ \,
\eea
In this set of basis,
we obtain the following two $4 \times 4$ 
determinant equations for the even and odd sectors of $\sigma_{vx}$,
respectively, as
\bea
\text{det} 
\left(U_{+,{e}}\tilde{\Psi}^{+}_{e},  U_{-,{e}}\tilde{\Psi}^{-}_{e}\right) = 0, 
\nn \\
\text{det} 
\left(U_{+,{o}}\tilde{\Psi}^{+}_{o},  U_{-,{o}}\tilde{\Psi}^{-}_{o}\right) = 0, 
\eea
in which $U_{\pm,e}$, $U_{\pm,o}$ are $4\times4$ matrices, 
and $\tilde{\Psi}^{\pm}_{e} = 
(\phi^{\pm}_{e,\frac{3}{2}},\phi^{\pm}_{e,\frac{1}{2}})$, 
$\tilde{\Psi}^{\pm}_{o} = 
(\phi^{\pm}_{o,\frac{3}{2}},
\phi^{\pm}_{o,\frac{1}{2}})$ are $4\times2$ ones.
The surface midgap state spectra displayed in the main text are 
solved from this set of equations which are 
fourth order algebraic equations of $\epsilon^2$.



\end{document}